\documentclass[conference,compsoc]{IEEEtran}
\ifCLASSOPTIONcompsoc
\else
\fi
\ifCLASSINFOpdf
\else
\fi
\usepackage[hyphens]{url}

\usepackage{tikz}
\usepackage{caption}
\usepackage{subcaption}
\usepackage[margin=1in]{geometry}
\usepackage{booktabs}
\usepackage{array,multirow, makecell}  
\usepackage{wasysym}                 
\usepackage{xcolor}    
\usepackage{graphicx} 
\usepackage{natbib}
\usepackage{enumitem}

\usepackage{color}
\usepackage{rotating}
\usepackage{tabularray}

\begin{document}
%
\title{Generative Propaganda \\ (Working Paper)
}

\author{
\IEEEauthorblockN{Madeleine I. G. Daepp}
\IEEEauthorblockA{\small Microsoft Research\\
Redmond, WA, USA\\
mdaepp@microsoft.com}
\and
\IEEEauthorblockN{Alejandro Cuevas}
\IEEEauthorblockA{\small Princeton University\\
Princeton, NJ, USA}
\and
\IEEEauthorblockN{Robert Osazuwa Ness}
\IEEEauthorblockA{\small Microsoft Research\\
Redmond, WA, USA}
\and
\IEEEauthorblockN{Vickie Yu-Ping Wang}
\IEEEauthorblockA{\small Independent Researcher\\
New York, NY, USA}
\and
\IEEEauthorblockN{Bharat Kumar Nayak}
\IEEEauthorblockA{\small Independent Researcher\\
Ranchi, Jharkhand, India}
\and
\IEEEauthorblockN{Dibyendu Mishra}
\IEEEauthorblockA{\small Cornell University\\
Ithaca, NY, USA}
\and
\IEEEauthorblockN{Ti-Chung Cheng}
\IEEEauthorblockA{\small University of Illinois\\Urbana-Champaign \\
Urbana, IL, USA}
\and
\IEEEauthorblockN{Shaily Desai}
\IEEEauthorblockA{\small Imperial College London\\
London, UK}
\and 
\IEEEauthorblockN{Joyojeet Pal}
\IEEEauthorblockA{\small University of Michigan\\
Ann Arbor, MI, USA}
}


\maketitle

\begin{abstract}
  Generative propaganda is the use of generative artificial intelligence (AI) to shape public opinion. To characterize its use in real-world settings, we conducted interviews with defenders (e.g. factcheckers, journalists, officials) in Taiwan and creators (e.g. influencers, political consultants, advertisers) as well as defenders in India, centering two places characterized by high levels of online propaganda. The term ``deepfakes'', we find, exerts outsized discursive power in shaping defenders' expectations of misuse and, in turn, the interventions that are prioritized. To better characterize the space of generative propaganda, we develop a taxonomy that distinguishes between obvious versus hidden and promotional versus derogatory use. Deception was neither the main driver nor the main impact vector of AI's use; instead, Indian creators sought to persuade rather than to deceive, often making AI's use obvious in order to reduce legal and reputational risks, while Taiwan's defenders saw deception as a subset of broader efforts to distort the prevalence of strategic narratives online. AI was useful and used, however, in producing efficiency gains in communicating across languages and modes, and in evading human and algorithmic detection. Security researchers should reconsider threat models to clearly differentiate deepfakes from promotional and obvious uses, to complement and bolster the social factors that constrain misuse by internal actors, and to counter efficiency gains globally. 
\end{abstract}
\section{Introduction}\label{sec:intro}

Generative AI can now produce text, audio, and images that most people cannot distinguish from real content~\citep{frank2024representative, diel2024human}. Major breakthroughs occurred just ahead of 2024---dubbed ``the biggest election year in history"---when more than half the world's population participated in major elections~\citep{economistelections}.   Elections are consistently targeted by information manipulation and interference campaigns~\citep{bradshaw2019global}, leading to widespread predictions that 2024 would see a ``deepfake deluge,'' an inundation of malicious and misleading AI-generated depictions of people or events~\citep{deepfakedelugetimes, deepfakedelugeventurebeat, wefrisks2024}. 

The potential of generative AI to produce deepfakes is unsettling. There is a long-standing history of sudden and massive reversals in political fortunes based on single pieces of information~\citep{reyes2006swift, twomey2023deepfake}. High-profile cases of viral AI-generated content have amplified concerns, including a 2019 video of former U.S. President Barack Obama (specifically intended to highlight the technology's risks in politics) \citep{silverman2018-spot-deepfake} and a 2023 image of a bombed Pentagon that catalyzed a stock-market sell-off (though the market quickly recovered)~\citep{nyt-2023-ai-spoof-markets}. In the global 2024 elections cycle, however, the use of generative AI to convince people of candidate interactions and events that did not, in fact, occur, was limited relative to researchers' expectations~\citep{schneier2024review}.\footnote{There are several important exceptions in which fraudulent deepfakes were both used and considered impactful \citep{meaker2023slovakias, indiatoday2024sule}; for a detailed review of AI incidents in 2024, see~\citet{trauthig2025}.} While there is little doubt that deepfakes in and of themselves are concerning, this article shows that the focus on antagonistic deception obscures a much broader space of generative AI's use and misuse in political communications.

Through in-depth study of generative AI's use in Taiwan and India, two places with national elections in 2024 and subject to high levels of computational propaganda~\citep{vdem2019, wefrisks2024}, we show that observed uses were a superset of those that journalists, fact-checkers, and other front-lines defenders expected. Developing a taxonomy that differentiates obvious versus hidden and promotional versus derogatory use, we find that influencers, political consultants, and other creators make AI's use obvious or promotional in order to minimize potential legal and reputational costs; social context further serves as a key bottleneck on whether content resonates or attains reach. Notably, these explanations contrast with common technologically determinist explanations that center limitations on quality, cost, and distribution as defining constraints on deepfakes' use~\citep{simon2023misinformation, kapoor2024, kapoor2023}. 

We further provide evidence of additional adversarial uses beyond deceptive representations. In particular, creators and defenders describe the use of ``AIPasta''---the AI-based perturbation of online communications to introduce authentic-seeming variation~\citep{dash2025persuasive}---and ``Precision Propaganda'', or AI's use to tailor messaging to narrow target audiences. The primary risk, we find, lies not in photorealism but in new efficiency gains—--lower detectability, multilingual reach, and multimodal output—--that enable actors to scale information manipulation across places and platforms, and for which existing social defenses provide little restraint.

Our work broadens the evidence of AI's use beyond representational deepfakes to highlight growing evidence of its value both for obvious and promotional representations as well as for adversarial efforts to attain narrative dominance. To respond effectively, security researchers should account for uses and misuses beyond deepfakes, prioritize interventions that match the scale of AI's efficiency gains, and recognize that social and technical defenses will serve as necessary complements in robust defensive systems.

Our article makes four contributions. 

\begin{enumerate}
    \item We collected 64 hours of interviews with 72 participants across two of the places most affected by computational propaganda globally, documenting front-line perspectives on the use and misuse of a major emerging technology.
    \item We provide a taxonomy to differentiate adversarial deepfakes from other representational uses, showing how persuasion and distortion motivated observed patterns motivated applications beyond antagonistic deception.
    \item We document observed threat actors and their constraints, including legal fears, reputational costs, and social context---factors that differ from leading prior theories.
    \item We provide detailed insights to inform threat models and interventions, highlighting the importance of tailoring mitigations to the sociotechnical context of a given place.
\end{enumerate}

\section{Literature Review}\label{sec:litrev}

Our study is rooted in three related bodies of work: (1) studies monitoring global information disorder; (2) interdisciplinary examinations of generative AI's misuse in digital communications, particularly via deepfakes; and (3) qualitative research developing human-centered frameworks for threat modeling in the security literature.

\subsection{Global Information Disorder} 

Global information disorder refers to the widespread use of computational propaganda---political messaging produced and amplified via automated, algorithmic, and surveillance technologies~\citep{bradshaw2019global}---to manipulate digital public spheres, the online spaces where public opinions are formed~\citep{habermas2023new}. Although early research focused on misinformation and disinformation----unintentional and intentional deception, respectively~\citep{jack2017lexicon}---researchers increasingly recognize that antagonistic falsehoods are just one subset of the broader class of efforts to manipulate collective sensemaking processes online~\citep{starbird2024facts}. The set of countries affected has expanded rapidly, more than doubling from 28 in 2017 to 70 in 2019~\citep{bradshaw2019global}; effectively, all democracies now confront some form of online information disorder. Despite conflicting research on the efficacy of such efforts in effectively deceiving targets~\citep{budak2024misunderstanding}, there is now strong evidence to show that existing schisms in society are not just amplified, but weaponized in digital public spheres~\citep{akbar2021misinformation, pawelec2022deepfakes, king2017chinese, benkler2018network}.

Advances in generative AI have reinforced the existing concerns posed by these research, with experts naming its use in disinformation as a leading global risk~\citep{wefrisks2024}. Particular concerns focus on how the rapid spread and accessibility of these technologies can ``democratize'' information manipulation beyond governments to participatory or volunteer actors~\citep{woolley2018computational, starbird2019disinformation}. Moreover, because such fears are grounded in earlier literature on deception (mis- and disinformation), defensive research has disproportionately taken the form of an arms race around deepfake detection~\citep{heidari2024deepfake}. In short, the overwhelming focus of work on generative AI's use in online communications has been on adversarial and non-consensual misuse, with limited attention to the possibility of willful and disclosed use or applications beyond deception.

\subsection{Deepfakes in the Security Literature}

There are three major lines of research to characterize deepfakes in the security literature. First, several studies have analyzed online message boards to understand creators' motivations. This literature has tended to focus on non-consensual intimate imagery~\citep{han2025characterizing, timmerman2023studying, gibson2025analyzing} and fraud~\citep{richet2022cybercriminal}, for which there are large and accessible online communities; no comparable online, public community exists, to our knowledge, for political uses. A second line of research looks at the capabilities of generative AI tooling to predict likely misuse~\citep{horvitz2022horizon, mehta2023can, frank2024representative}. Social factors, however, shape the conditions under which such capabilities are actually used \citep{mackenzie1999social}, highlighting the need to investigate how and why real-world use diverges from these predictions. A lack of attention to the ``mutual shaping'' of society and technology has also led to a "techno-chauvinistic" narrative of deepfakes as a historically unprecedented epistemic threat whose solutions only lie in technological defenses~\citep{habgood2023deepfakes, paris2019deepfakes}. Highlighting the similarity of past ``moral panics'' around the deceptive capabilities of earlier media technologies, these theorists remind us to pay attention to the social norms and contexts in which deepfakes are produced and disseminated, and find ways to regulate and de-incentivize these spheres.  

Finally, a third line of work relies on observations of misuse, including incident databases~\citep{trauthig2025} and media reports~\citep{kapoor2024}, to track AI's misuse in global democracies. A major challenge in this literature is the contestation of the term ``deepfakes''. Journalists and researchers use the term to describe non-consensual images of people~\citep{nytimes2025riseAIinfluencer}, any believable but AI-generated depictions of people or events~\citep{horvitz2022horizon, widder2022limits, kietzmann2020deepfakes, frank2024representative}, any use of AI in representing people~\citep{meng2024ava, le2025sok}, or all content created with generative AI~\citep{han2025characterizing, gamage2022deepfakes, abdullah2024analysis}. Given this confusion, trackers include consensual, benign, and even positive in addition to malicious examples~\citep{trauthig2025, kapoor2024}---creating a challenge for security researchers, who need to understand the conditions under which generative AI's misuse poses a threat. A contribution of our study is thus to offer a taxonomy to aid in the differentiation of adversarial AI-generated representations from other uses, and to document additional misuses that may currently be evading journalistic or computational discovery.

\subsection{Informing Threat Models Through Qualitative Inquiry}

Our work also contributes to a line of research in the security literature that interviews people to understand human-centered threat models and how these models inform technical interventions~\citep{usman2025security}. These include studies of the security and privacy needs of understudied groups such as journalists~\citep{mcgregor2015investigating}, humanitarian organizations~\citep{le2018enforcing}, immigrants~\citep{usman2025security, tran2024security} and survivors of intimate partner violence~\citep{matthews2017security,freed2019my}; research on why user behaviors diverge from best practices~\citep{pearman2019people, munyendo2023eighty, ray2021older, redmiles2016think}; and work to understand people's perceptions of different security and privacy threats~\citep{munyendo2025you, cao2024understanding} including of generative AI~\citep{yu2025exploring, pena2025journalists}. These studies highlight how non-technical factors (e.g., laws, organizational practices, culture, etc.) may be in tension with technical assumptions and ultimately undermine the security of systems. We contribute to this literature by providing qualitative evidence on how frontlines defenders---journalists, factcheckers, researchers, officials---perceived and responded to generative AI risks during a major global elections year, and surfacing contextual and social factors to inform security researchers' threat models and interventions.

\section{Methods}

This research draws on two case studies of generative AI's use and misuse. Our work is grounded in a social shaping perspective, following~\citet{mackenzie1999social}, that seeks to understand how technical and social systems inform each other's development and use. That is, we pay active attention to its "mutual shaping" with social, economic and political factors through situated and contingent processes. In subsequent sections, we develop our taxonomy by distinguishing not just the technical forms but also the socioeconomic contexts of persuasion, reputation, and efficiency, the legal landscape, and local media ecologies. 

\subsection{Settings and Sampling Approach}

We selected Taiwan and India as case studies based on their high prevalence of information disorder~\citep{vdem2019, wefrisks2024}, the diversity of their defensive ecosystems~\citep{barron2025taiwans, travelli2024nyt}, and the fact that each was set to hold major national elections in 2024~\citep{economistelections} Our general strategy in both cases was to first purposively sample participants through expert guidance and then to snowball the sample until we reached saturation. \\

\textbf{Taiwan.} We conducted fieldwork in Taiwan from December 20, 2023 to January 19, 2024, coinciding with the election on January 13. We identified a preliminary set of interviewees based on guidance from Taiwan Studies experts as well as on prior research on Taiwan's civic technology and media ecosystems~\citep{lee2020nobody}. We sought to interview people we term ``defenders'', or members of organizations focused on tracking and combating information manipulation, including fact-checkers, civic technologists, researchers, and journalists. We used snowball sampling, which expanded our set of participants to stakeholders across national security, cybersecurity, and private industry in addition to civil society. Most of the interviews were held in Taipei; the team also traveled to Hsinchu and Taichung to meet participants. Our research team included a professional interpreter, and interviewees were given the option to have the interview conducted in either Chinese or English. We also conducted participant observation in public events and meetings; when in small-group settings, we read a short statement to participants detailing our research and data subjects' rights. 

\textbf{India.} We conducted fieldwork in India between June 3 and June 21, 2024, spending a week each in Bangalore, Delhi, and Mumbai. India's elections occurred in phases between April 19 and June 1, with results reported on June 4, and thus our interviews were generally conducted after the event and focused broadly on AI and digital communications rather than narrowly on political use. We again sought to interview defenders including fact-checkers, journalists, and lawyers; we further expanded our sampling to include ``creators,'' or people involved in the creation of online content, such as political consultants, social media influencers, technologists, and advertisers. We leveraged an initial set of contacts from research team members' prior work in the Indian context, again expanding our set of interviewees via snowball sampling. Although interviews were generally conducted in English, at least one multilingual team member attended all but 3 interviews in India to ensure that participants had the option to switch languages. 

\subsection{Interviews and Analysis}

We conducted semi-structured interviews, allowing lines of questioning to evolve over the course of each study. In each interview, we focused on the participant's work before asking about generative AI's impacts, first through general questions around changes in the past year and then with probes specific to real-world cases of AI's use or AI-powered tooling. Interviews were recorded using Microsoft Teams and transcribed and analyzed with the Hey Marvin qualitative research software. Interviews were intended to take approximately one hour with one subject, but interviews with organizations frequently included multiple participants and ran for several hours. Our final data set included 55 unique interviews with 72 participants (35 and 37 participants in Taiwan and India respectively), comprising 64 hours of data in total. All participants provided verbal consent, and interviewees were given the option to determine how they wished to be identified (by name/title, role, organization, or anonymous).

After each interview, meeting, and event, we wrote short memos detailing the content and relevant themes. When it was not possible to write a memo immediately or when the lead researcher was not present, the memo was compiled later from notes, discussion, or upon reviewing the recording. We further compiled these memos into weekly reports---based on the interviews as well as attendance at events and meetings and ongoing observations of online artifacts---that we circulated for feedback from experts across security research, political science, and generative AI.

The lead researcher coded interviews through a two-phase, abductive coding approach, iterating between inductive theory-building and deductive application of existing frameworks~\citep{saldana2011fundamentals}. In the first phase, she reviewed a sample (20\%) of the interviews from each setting (Taiwan n = 4, India n = 7), using rapid open-coding to develop an initial set of over 100 codes. She then reviewed this set of codes, as well as memos and research reports, to construct a mind map that constrained the codebook to five major themes. After reviewing the set of codes and themes with other members of the research team, she applied the codebook to the complete corpus of interviews. Reliance on a single researcher to conduct the coding, although burdensome, had the benefit of ensuring internal consistency across the application of codes and in recognizing the role of the researcher as a key data instrument~\citep{geertz2017interpretation}. However, the lead researcher is neither Taiwanese nor Indian, and thus consulted with members of the research team who do hold those identities for feedback on the analyses and findings.

\subsection{Positionality and Ethics}

We recognize that conducting research across two places with different contexts and languages is fraught with important questions regarding the impact of power differentials, linguistic barriers, and contextual knowledge (or lack thereof) on the validity of the results. We took three measures to recognize and mitigate these challenges. First, we recruited researchers from each studied geography to join the research team in order to foster reflexivity and ensure broader perspectives informed data collection and analysis~\citep{barry1999using}. The coauthors hold a range of relevant identities, and have firsthand experience working as fact-checkers (BN), journalists (VW, BN), and civic technologists (DM, TCC) in the contexts studied. Second, we provided interpretation support and invited participants to choose their preferred language for the interview; however, language has status connotations in both Taiwan and India and thus participants may nevertheless have felt pressure to participate in English. Third, we validated our representations of participants by asking participants explicitly how they wished to be identified (by name, title, role, organization, or anonymously) and by asking participants to review quotations in advance of publication as a form of member checking~\citep{small2022qualitative, harvey2015beyond, guba1981criteria}. While it is standard practice to anonymize qualitative research participants in security and privacy research, making anonymity the default risks amplifying power asymmetries by preventing recognition of local expertise and reducing the agency of participants who \textit{want} attribution for their insights~\citep{tilley2011end}. We also note that compensation differed across the two contexts: In Taiwan, participants generally operated as representatives of professional organizations, and thus were not compensated; In India, where we generally interviewed just one or two participants at once, participants were compensated between \$100 and \$400, rates assessed based on appropriate payments for professional digital consultants. Both studies were reviewed and approved by the Microsoft Research Institutional Review Board; the study in India was also reviewed and approved by the University of Michigan Institutional Review Board.

\begin{table*}
\centering
  \caption{Conceptual taxonomy of generative AI's use in representations. }\label{tab:taxonomy}
    \begin{tblr}{
        width = \linewidth,
        colspec = {Q[10]Q[10]Q[500]Q[500]},
        row{2} = {c},
        column{2} = {valign=m, halign=r, rightsep=8pt},
        cell{1}{1} = {c=2,r=2}{0.02\linewidth},
        cell{1}{3} = {c=2}{c},
        cell{3}{1} = {r=2}{c},
        vline{3-4} = {3-4}{},
        hline{1,5} = {-}{0.08em},
        hline{3} = {3-4}{},
        hline{4} = {3-4}{},
    }
    &           & \textbf{Use of Generative AI}           &           \\
    &           & Obvious           & Hidden           \\
    \begin{sideways} \textbf{Mode of Portrayal}\end{sideways} 
        & \rotatebox[origin=r]{90}{Promotional\hspace{0.5em}} 
        & {\textbf{Soft Fakes:}\\Humorous or laudatory AI-generated representations.
            \\\\\textit{Example:}~voice clones of candidates singing}           
        & {\textbf{Auth Fakes:}\\ Serious AI-generated representations authorized by the candidates they portrayed.
            \\\\\textit{Example}: speech videos dubbed across languages}\\
        & \rotatebox[origin=r]{90}{Derogatory} 
        & {\textbf{Deep Roasts:}\\Satirical representations in which candidates were mocked or humorously denigrated.
            \\\\\textit{Example}: social media filter replacing user face with angry, chiding candidate} 
        & {\textbf{Deep Fakes:}\\Serious representations of candidate activities or events that did not actually occur.
            \\\\\textit{Example}:~speeches altered to include controversial statements} 
\end{tblr}
\end{table*}

\section{Observed Uses of Generative AI}\label{sec:taxonomy}

Both Taiwanese and Indian defenders described expecting the predominant misuse of generative AI to be for deceptive, adversarial representations of candidates and events. But observations of such uses were infrequent relative to salient uses overall. The technology reporter Nilesh Christopher described the Indian landscape:
\begin{quote}
    \textit{``Basically everybody thought deepfakes were going to have this scorched earth future, but it's effectively being co-opted only for two purposes. One is voter outreach and another one is political memes, for shits and giggles.''}
\end{quote}
While relatively few deepfakes were observed, usage of generative AI overall comprised a superset of the uses defenders expected. This mismatch between expectations and practice highlights the need to better parse how generative AI was actually used in political communications. 

\subsection{A Taxonomy of Use}

We reviewed each instance in which participants used the term ``deepfake'' as well as all cases in which generative AI was used. Below, we name and describe the major types of observed uses. \\

\noindent \textbf{Representational Use} We first review each case in which AI was used to represent candidates. We construct a conceptual taxonomy for these cases, presented in Table~\ref{tab:taxonomy}, grounded in two key distinctions. First, we look at  whether AI's use was obvious or hidden. Obviousness was sometimes made explicit, with creators including watermarks and labels; more frequently, it was implicitly evident via hyperrealistic or cartoonish aesthetics. Second, we assess whether the depiction was promotional or derogatory. Despite the common expectation that generative AI would primarily be used to attack candidates, many depictions lauded or lifted up the candidates they portrayed.

These distinctions surface four major categories of representational use:
\begin{itemize}[leftmargin=*]
    \item \textbf{Soft Fakes (Promotional / Obvious).} In songs, images, and videos that proliferated across Indian social media, influencers used religious iconography or hypermasculine features in portraying the incumbent prime minister, for example, and superimposed him onto a video of a dancing rapper.  \citet{chowdhury2024AIConnectII5} coined the term ``soft fakes'' to reflect how such content softens the image of portrayed candidates.
    \item \textbf{Auth Fakes (Promotional / Hidden).} A number of candidates in India authorized AI-generated portrayals of themselves. Indian campaign teams commissioned holy wishes in the candidate's voice, for example, and dubbed speeches across different languages. We call such usage ``auth fakes'' in recognition that the representation was authorized by its subject.
    \item \textbf{Deep Roasts (Derogatory / Obvious).} Users also produced parodies (e.g. a face-swap of two opposition leaders into a Bollywood scene in which actors bemoan their many failures, and a filter that enabled Taiwanese users of TikTok to portray a presidential candidate as angry and off-putting). We denote these uses ``deep roasts'' in recognition that creators were ``roasting''---a prevalent form of insult comedy---or mocking the candidates they opposed. 
    \item \textbf{Deep Fakes (Derogatory / Hidden).} By reserving the term ``deepfake'' for those cases when the portrayal is derogatory and generative AI's use is hidden, we are able to distinguish a small subset of cases that clearly reflect deceptive and adversarial usage.
\end{itemize}

\begin{figure*}[htbp]
  \centering
  \includegraphics[width=1\textwidth]{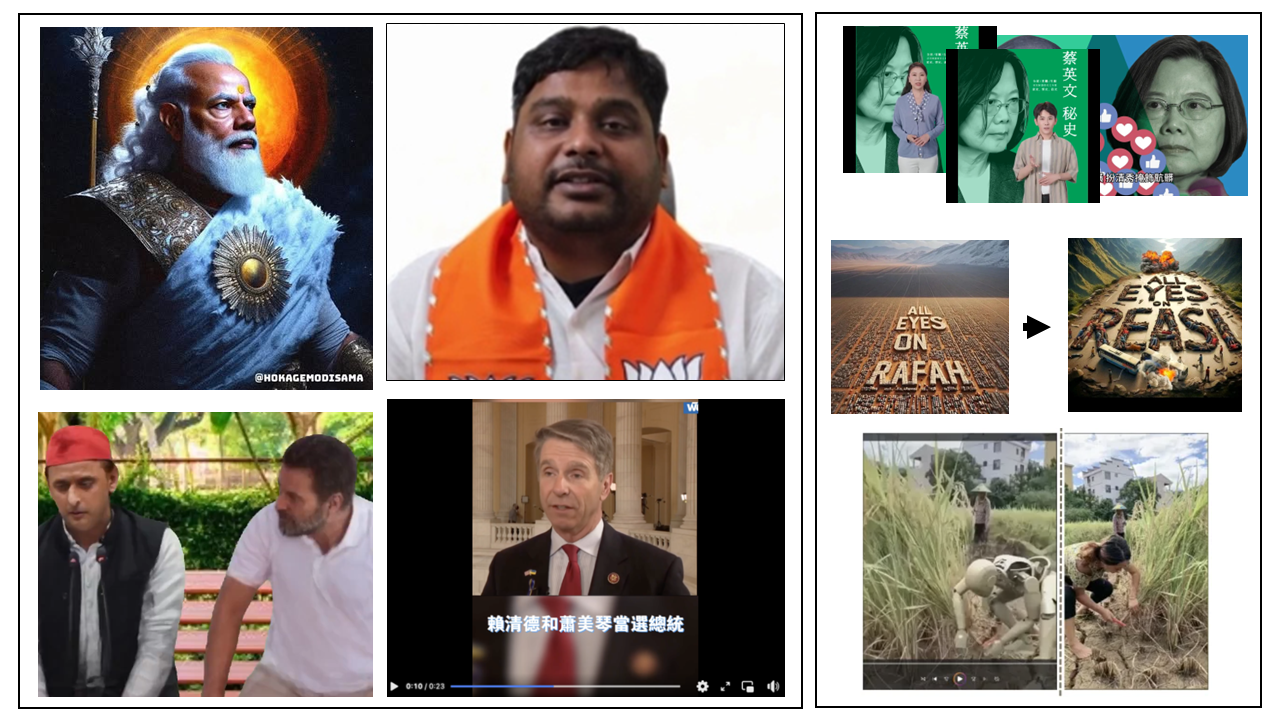}
  \caption{Examples of generative AI's usage. The left panel includes, clockwise from top left, a soft fake applying religious iconography to incumbent prime minister Narendra Modi~\citep{christopher2024-instagram-ai-modi}; an auth fake in which a politician's speech is dubbed across languages~\citep{dau2024-rajasthan-avatar-tamil}; a deepfake in which a U.S. congressman's speech is altered~\citep{li2024-seeing-not-believing-ii}; and a deep roast in which two opposition candidates are face-swapped into a Bollywood scene in which actors bemoan their life's failures~\citep{x_soldiersaffron7_clip}. The right panel includes from top to bottom: AIPasta (YouTube videos of AI-generated news anchors that proliferated just before Taiwan's election)~\citep{tsai-chin2024-secret-history}; Precision Propaganda, in which a globally viral AI image is altered to emphasize events of concern to Indian audiences~\citep{adnal2024reasi}; and AI Slop (videos in which AI was used to swap human and robot farmers)~\citep{tfc_10288_robot_farming_2024}.} 
  \label{fig:examples}
\end{figure*}

\noindent \textbf{Usage Beyond Representations.} A number of important use cases of generative AI did not fit within the ``deepfakes'' categorization or the broader space of use for portrayals of candidates. These included:
\begin{itemize}[leftmargin=*]
    \item \textbf{AIPasta.} The use of generative AI to introduce artificial variation into social media posts---so-named because it represents an update to previous CopyPasta methods in which actors copy and paste content in order to achieve scale~\citep{dash2025persuasive}.\footnote{Or: throw lots of content at the wall and see what sticks.}
    \item \textbf{Precision Propaganda.} The use of AI tooling to micro-target or tailor the content of media to the identities of a targeted audience (e.g. tailoring videos, prompted from b-roll and stock imagery, to better align with the caste and occupation of the targeted audience in India).\footnote{We draw the term from the concept of precision medicine, the tailoring of drugs based on the genetics and physiology of the patient, because---at its most extreme---actors target content based on an opinion model of an individual, as described in~\citet{goldsteinBenson2025AIpropaganda}.}
    \item \textbf{AI Slop.} Low-quality or absurdist AI-generated content, observed in both places, and which journalists have documented across different geographies across the world~\citep{hoffman2024first}.
\end{itemize}

\noindent \textbf{Analytic Usage.} Participants in India also described applications of AI in data analysis and media monitoring. Political consultants emphasized the advantages of AI-driven phone survey tools in attaining scale, speed, and consistency, as well as the role of AI in speechwriting, media and social media monitoring, and crowd sentiment analysis. Although these applications shape the decisions of both campaign and external influence operations, their direct impact on public discourse is relatively limited. For this reason, our focus in this article remains on usage in content creation and sharing, rather than on analytics.

\subsection{From Deepfakes to Generative Propaganda}

To understand the importance of ``deepfakes'' as a term in shaping the expectations and, in turn, interventions that many defenders prioritized, we introduce the concept of discursive power. \textbf{Discursive power} denotes the extent to which a linguistic convention or symbolization conveys a particular mental model of a problem, in turn determining who is empowered to solve it and how~\citep{reed2013power}. Consistent with the discursive power of deepfakes as a concept, concerns around AI's use in deception shaped major preventative efforts in both places that we studied. In Taiwan, prominent fact-checking and media literacy organizations published guidance and held trainings to help the public identify the signs of generative AI's use for deception; in India, media outlets and fact-checkers formed coalitions with dedicated deepfake tracking and analysis units; and in both places, government or civil society actors built or procured detection tooling. Defenders' efforts helped mitigate adversarial deepfakes. However, an overemphasis on mitigating deceptive use risks overlooking substantial additional uses. That is, we observed many cases in which AI was used for non-deceptive or non-representational content that aimed to shape recipients' perceptions of people and events---a classical definition of propaganda~\citep{ellul1965propaganda}. We thus call this broader space of use ``generative propaganda.''

\section{The Usefulness of Generative AI}\label{sec:motivations}

Whether and how generative AI was used was shaped as much by the motivations of key actors as by the technology's capabilities. Understanding efficiency gains beyond deceptiveness, moreover, is key to further expanding the space of misuse. 

\subsection{Motivations} 

We first examine the motivations that surfaced, among our participants, for AI's use in digital communications. We note that deception, in these contexts, was a means to an end, with primary motivations of persuasion and distortion helping to explain the proliferation of uses beyond adversarial deepfakes.\\

\noindent \textbf{Persuasion.} A primary goal of Indian creators was to \textit{persuade} audiences to see candidates or events in a particular light. Creators using AI to represent candidates, particularly through soft fakes and deep roasts, sought to set the frame through which viewers understood the candidate---recognizing that frames, once set, shape how new evidence is interpreted~\citep{starbird2024facts}. Such persuasion can entail deception, but it is not required; indeed, making AI's use obvious did not compromise its persuasive capabilities. Describing the creation of a soft fake of a political candidate, for example, the political consultant Ritwick Shrivastav explained:
\begin{quote}
    \textit{``We purposefully made it more cartoony, so that it's clear to the person who's seeing this image that this is not a real image. This is an AI-generated image. But the people who understand that narrative do not care whether the image is real. The messaging that you wanted to deliver, you have delivered that.''} 
\end{quote}
Making a representation obvious reduced its deceptive capability (and its social risk), but not its persuasiveness (Section~\ref{sec:threatmodel}). \\

\noindent \textbf{Distortion} In Taiwan, \textit{distortion}---ongoing efforts to distract from some narratives and amplify others---was a major and ongoing challenge. Flurries of AIPasta after a mistranslated government alert (reporting a missile launch rather than a satellite flyover), for example, showed the technology's use in efforts to set perceptions of emerging events~\citep{daepp_ness_2024_video_will_kill_truth}. Describing the rapid and scaled proliferation of online narratives after or around major events, Taiwan's digital minister Audrey Tang said:
\begin{quote}
     \textit{``So it's crowding out. It's a DDoS. It's a denial of service on the democratic language, which is always so limited before the election anyway.''}
\end{quote}
In Taiwan as in India, the intention was not always to lie; rather, the disinformation researcher Tim Niven described, ``what can be treated with the frame of factuality is a subset of the content we see" but ``narrative is everywhere.'' (See also \citep{niven2023evolution}.) Consistent with prior research on the importance of ``strategic distraction'' as an information manipulation tactic~\citep{king2017chinese}, it was generative AI's use in efforts to attain narrative dominance through the seeding, amplification of, or distraction from strategic narratives---rather than the technology's isolated deceptive capabilities---that posed the most serious concern. \\

\subsection{Efficiency Gains} For the purposes of persuasion and distortion, we observed three key efficiency gains that generative AI offered to creators: reduced detectability, multimodality, and multilingualism. The increases in these capabilities represented a major change over the previous state of the art (in contrast with photorealism, which could often be obtained with existing ``cheapfake'' methods~\citep{paris2019deepfakes, kapoor2024}). We detail each of these efficiency gains and their implications below. \\ 

\noindent \textbf{Reduced Detectability.} First, AIPasta and Precision Propaganda evaded detection processes by inducing artificial variation and targeting specific groups or individuals, respectively. Describing AIPasta's successful use, a social media influencer and political communications expert explained: 
\begin{quote}
    Influencer: \textit{``If you ask someone to write those captions, he may not get that kind of variety of words or ideas that they are getting from ChatGPT.''} \\
    Interviewer: \textit{``To get a hashtag trending, for example, does that variety matter?''} \\
    Influencer: \textit{``It does. Because in between there were these fact checkers and news watchers, they do catch that there is some spamming going on Twitter. If you don't have variety in your captions and content, then people will point it out that, okay, this is a sponsored thing.''} \\
\end{quote}
Operators continued to pay humans to scale the posting of content in India, where such labor was cheap, but they used generative AI to induce artificial variation---with the explicit aim of evading both human and algorithmic defenses. Evading human detection, in particular, has the potential to make AIPasta more effective as a tool for distortion: experimental research shows that people are significantly more likely to see AIPasta as reflective of a true social consensus, relative to prior CopyPasta methods~\citep{dash2025persuasive}. \\

\noindent \textbf{Multimodality}. In both Taiwan and India, there was clear evidence consistent with the usefulness of generative AI in enabling adversarial actors to expand both persuasive and distortive operations across modes. A political consultant in India, for example, walked us through the use of prompting to turn what might previously have been a message conveyed via static text or imagery into a video; and Taiwanese media literacy and fact-checking organizations highlighted the extent to which they were seeing content farms expand from text-based operations into video-based content. Such capabilities were particularly important, creators noted, given the increasing importance of short-form video platforms in people's social media diets. Video, notably, was harder for defenders to monitor or counter relative to text- and image-based social media---most video platforms have no reverse search tooling for defenders seeking to track a video reported through a tipline back to its source. Moreover, few of the defenders we interviewed were active on short-form media platforms, with most fact-checking organizations continuing to provide their findings via text articles rather than videos. \\

\noindent \textbf{Multilingualism.} Finally, generative AI enhanced the ability of creators to operate in languages different from their own. The political consultant Ritwick Shrivastav described interacting with local staff he had hired:
\begin{quote}
    \textit{They were using ChatGPT on their mobile to translate the local language to English, and they were sending me that. It was broken English, but at least it was legible and understandable for me, which is a huge plus. So now I have the ability, being a North Indian, to work in any part of India, because I know that we'll be able to figure out what the locals are saying.}
\end{quote}
AI's capabilities remain constrained for low-resource languages like many of those spoken across India, but it was already expanding the ability of a small team to scale to new places. That similar changes were at play in Taiwan's information ecosystem was evident in a change in behavior of troll groups. Describing why he considered the behavior of social media accounts he had identified as likely inauthentic, Taiwan AI Labs founder Ethan Tu explained, 
\begin{quote}
    \textit{``Taiwanese people were not fighting with each other on PTT [an online forum he had founded] and at the same time fighting with each other in English in the United States.'' }
\end{quote}
He was seeing accounts that produced high-quality content in multiple languages depending on the target and audience---but \textit{real} Taiwanese people would not have been able to translate so rapidly and well across languages. In Taiwan, in particular, mistakes in the conversion of Simplified to Traditional Chinese have long been a key tell that defenders used to identify influence operations; AI helps adversaries overcome this defense. As Tang pointed out, the power of generative AI was that distortion ``scales across cultures.'' 

\subsection{Harms}

The harms associated with efficiency gains in AI's misuse in political communications differed between the two contexts studied. In both contexts, the perceived harms associated with deepfakes and deception were challenges to election integrity. More generally, for Taiwanese defenders, major harms associated with AI's misuse and information disorder included challenges to self-determination and political polarization. For Indian defenders, in addition to concerns about polarization, the potential of content to reinforce communal divisions, catalyzing violence across religious and caste backgrounds, also surfaced as a major perceived harm.\footnote{Both places shared major additional concerns about AI's misuse beyond political communications, with both Taiwanese and Indian defenders noting increasing evidence of AI's misuse in scams and fraud; in India, AI's use in non-consensual intimate imagery for the purposes of ``sextortion'' or defrauding targets was also a prominent concern.} 
\section{Implications for Threat Models}\label{sec:threatmodel}

Our research can inform the work of researchers who develop threat models for this space. Throughout this paper, we have discussed the primary threats (concerns and harms) associated with generative propaganda. Following the human-centered threat modeling framework of \citet{usman2025security}, this section further discusses the main threat actors and examines the constraints that emerged through our research. We review how constraints differ between actors, with implications for the effectiveness of potential interventions. Social factors---legal fears, reputational risk, and social context---more than technical limitations, constrained misuses by internal actors and actors with consistent online identities, but not actors who were external or whose identities could be disposed and replaced. Technical barriers, such as model guardrails, were evident in the content produced by participatory actors, but did not encumber more technical actors. Through this analysis, we highlight that the key to understanding and predicting the impact of generative AI on information disorder is the efficiency gains, aligned with the persuasive and distortive motivations of key actors, afforded by generative AI rather than improved deceptive capabilities. Finally, we contrast our findings with existing theory to show the value of social factors in constraining deepfakes but not other forms of misuse.

\newcommand{\both}{\scalebox{1.1}{\CIRCLE}}      
\newcommand{\taiwan}{\scalebox{1.1}{\LEFTcircle}} 
\newcommand{\india}{\scalebox{1.1}{\RIGHTcircle}}
\newcommand{\neither}{\scalebox{1.1}{\Circle}}        

\renewcommand{\arraystretch}{1}
\newcolumntype{C}[1]{>{\centering\arraybackslash}p{#1}}  

\begin{table*}[htbp!]
\centering
\caption{Observed motivations and constraints across main threat actors.}\label{tab:threatmodel}
\begin{tabular}{lccccc}
\toprule
& \textbf{Nation State} & \textbf{Political Campaigns} & \textbf{Troll Groups} & \textbf{Influencers} & \textbf{Content Farms} \\
\midrule
\multicolumn{6}{l}{\textbf{Motivations}} \\ \midrule
Deception     & \taiwan     & \india    & \both     & \neither   & \both \\
Persuasion    & \taiwan     & \both     & \both     & \both     & \both \\
Distortion    & \taiwan     & \both     & \both     & \both     & \both \\
\addlinespace
\midrule
\multicolumn{6}{l}{\textbf{Constraints}} \\ \midrule
Reputational Risk  
                    & \neither   & \both   & \neither  & \both     & \india \\
Legal Ramifications     
                    & \neither   & \both   & \india    & \both     & \india \\
Social Context        & \taiwan   & \both   & \both     & \neither     & \both \\ 
\addlinespace
\bottomrule
\end{tabular}
\vspace{0.75em}

\textbf{Legend:} \both{} = observed in both contexts; %
\taiwan{} = Taiwan only; %
\india{} = India only; %
\neither{} = Not observed.
\end{table*}

\subsection{Threat Actors.} 

Throughout our interviews, five major actors emerged, as observed in Table~\ref{tab:threatmodel}: \textit{nation-states} (observed only for Taiwan); \textit{political campaigns} (more active in information operations in India); \textit{troll groups}, or networks of generally anonymous online actors who engage in coordinated inauthentic behavior; \textit{influencers}, content producers who generally use real or consistent pseudonymous accounts; and \textit{content farms}, actors who engage in the cultivation of online attention across platforms, building audiences from which they could receive advertising revenues or that they could sell as distribution channels for other actors.\\

\subsection{Constraints}

\noindent \textbf{Legal Risks.} Indian content creators were highly attuned to legal risks: A social media influencer and political communications expert explained how an arrest had made him more careful about the factuality of his posts; the founder of a social media marketing firm listed fears of legal retribution as among the top reasons he limited his postings of AI-generated content; and multiple founders of AI startups described rejecting ``unethical'' requests due to legal concerns. Several high-profile legal cases at the start of India's election season heightened the salience of legal fears. As Divyendra Singh Jadoun, founder of AI startup Polymath Synthetic Media Solutions, explained,
\begin{quote}
    \textit{``There was a deepfake of an Indian Bollywood actress Rashmika Mandana...Every news outlet was covering it, and our prime minister also said that deepfakes should be banned. So from that [deepfakes] were getting a lot of negative connotations. So I stopped creating. I thought my business was over. It was a time when I started getting these political clients, but I didn't want to take any political clients.''}
\end{quote} 
He and other technologists published an ethical compact, pledging to maintain transparency in AI's deployment, and he decided to include a visible watermark on the content he created. In India, where representational misuse could lead to defamation charges, making AI's use obvious was a strategy for reducing the legal risk to creators. Similar dynamics likely operated in Taiwan, where the Election and Recall Act made distributing or sharing deepfakes of candidates a crime punishable with prison time~\citep{tfc-2024-election-ai-audio-2023}; the legal risk was also made salient by ongoing reporting on the 2021 arrest of a YouTuber who sold pornographic deepfakes of politicians and other public figures~\citep{focusTaiwan2023deepfakeJail, shan2023deepfakeFraud}. Such legal measures could deter internal actors, particularly those with consistent and traceable public profiles such as local influencers and political campaigns, but they are more limited in constraining anonymous actors and do not deter actors operating external to the affected place. For example, content farms and troll groups that operate out of countries with few legal barriers, extradition treaties, or incentives to prosecute, will not be impacted by legal constraints.\\ 

\noindent \textbf{Reputational Economics.} Indian influencers also pointed to the high potential costs of deception, if discovered, for the value that their handles had accrued over time. Balram Vishwakarma, founder of social media marketing firm Scroll Back Studios, said,
\begin{quote}
    \textit{``Since I meet people every day and they know I run this page and everything, I can't do anything shitty. Even though I felt like when I was 21, like, can I make a quick buck? But then I realized I can't, because the industry is very small and if they find out that I'm doing something shady, they'll screw me up. Like, I won't get work afterward. Yeah. Reputation.''}
\end{quote}
He ran a successful Instagram page that had built economic value for him over time, enabling him to convert his influence into a social media marketing career---one he would lose if he were to cash in on his reach by spreading deceptive content.

Reputational risk was greatest when a specific identity (real or pseudonymous) was at stake. Faceless social media accounts, operated by troll groups or content farms, face more limited reputational risks. For a content farm, prior posts or additional contextual information like page origin could reduce the price other actors were willing to pay for their audiences; one actor, for example, described hiring an operation to comment on her behalf only to discover that the same accounts had previously posted content for the party she opposed, undermining the service's believability. Such content can generally be deleted before a sale, however. The lack of existing reputational markers in social media (e.g., indicators that the account had posted violating content in the past), then, and the ease with which accounts can change their identity (i.e., by changing their handles, names, deleting old content and re-uploading new different content) reduce potential reputational constraints for actors with disposable identities. \\

\noindent \textbf{Social Context.} People make sense of online content through collective sensemaking processes~\citep{starbird2024facts}. That is, shared social context shapes what content people pay attention to and how they interpret new evidence. For example, asked how she knew that a video of a U.S. senator supporting a Taiwanese party was fake, Eve Chiu, CEO of the Taiwan Factcheck Center, said 
\begin{quote}
    \textit{``By our journalistic expertise and our geopolitical knowledge. We know if a U.S. Congressperson says things like that, there will be an international outcry, in New York Times, in CNN, in BBC, not just on social media.''}
\end{quote} 
That is, defenders relied on social context before turning to technical tools for detection. Although the general public lacks comparable expertise, social context nevertheless remains an important factor shaping the collective sensemaking process and thus the impact of any particular piece of content. To land, then, deceptive portrayals needed to cohere with pre-existing or plausible perceptions and to align with existing or emerging attentional demands---a requirement that helps to explain why deepfakes usage was largely limited to strategic time windows (pre-election media blackouts; border skirmishes) in which perceptions were contested and evolving rapidly. The efficiency gains discussed in Section~\ref{sec:motivations}, however, offered direct reductions in the challenge of ensuring alignment with existing social context: troll groups could rapidly respond to new attentional demands as they emerged, and content farms could expand the variety of messages they tested in order to probe the attentional susceptibilities of different users.\\

\noindent \textbf{Technical and Distributional Barriers.} Existing theories for the relative dearth of deepfakes in recent elections posit that usage is constrained by technical factors (model capabilities, guardrails, and distribution bottlenecks) while impacts are limited by the saturation of misinformation demand~\citep{kapoor2023, kapoor2024, simon2023misinformation, cowen2023}. These factors can and do constrain participatory actors, but they are limited in their efficacy for constraining technically savvy or well-funded actors. In practice, technologists described bypassing model-level guardrails by using locally-hosted and open-source models, which in turn ensured the proliferation of software alternatives even for participatory operators. Political campaigns could bypass distribution bottlenecks, the platform-level mitigations that constrain the proliferation of content (e.g., group-size limits and bot-detection algorithms), by hiring influencers, paying content farms for posts, and messaging users directly on encrypted messaging platforms---that is, by tapping into an extensive grey market for distribution. \\ 

\subsection{Comparison to Prior Theory} The choice of whether to use generative AI can be understood as a function of the costs relative to the potential for impact. We found limited evidence for the efficacy of existing and theorized technical costs and barriers. Social costs, however---particularly legal and reputational blowback---emerged as important deterrents for adversarial deepfakes, albeit with varying impact depending on the threat actor; social context, moreover, pared potential impact. 

This simple cost-benefit analysis stands in stark contrast to existing theories for the relative dearth of deepfakes in recent elections, which posit that usage is constrained by technical factors (model capabilities, cost, and distribution bottlenecks) while impacts are limited by the saturation of misinformation demand~\citep{kapoor2023, kapoor2024, simon2023misinformation, cowen2023}. We note that these theories are alternately technologically determinist, describing technology as a driver of change and ascribing little efficacy to humans, or (in the case of demand saturation) social constructionist, presuming that humans determine technological change~\citep{baym2015personal}. 

Our work draws instead on social shaping theories of tech adoption, recognizing that sociotechnical systems are simultaneously shaped by both technical capabilities and social context, to identify the conditions under which reputational, legal, and social factors constrain fraudulent deepfakes in political contexts~\citep{mackenzie1999social}. In centering the role of social factors and cautioning against an overemphasis on deception, our work adds empirical evidence to bolster prior arguments, also grounded in social shaping theory, suggesting that techno-determinist arguments overstate the extent AI's deceptive capacity differs relative to prior technologies, and that social norms will emerge to keep misuse in check~\citep{habgood2023deepfakes, paris2019deepfakes}. 

We do, however, find that the efficiency gains entailed by AI's advances can and are being used to scale and expand persuasion and distortion across languages and platforms. Given a well-established, well-funded, and well-incentivized network of actors seeking to tilt public spheres in their candidates' favor and a novel technology with affordances suited to that work, the question is not whether but how generative AI was used. This is the question to which our work responds.

Our analysis suggests that persuasive, representational and distortive, non-representational misuses of AI, for which social costs are currently more limited, will proliferate---in contrast with representational deepfakes, which carried a high perceived cost for internal actors relative to their likelihood of success. Our work helps to explain why adversarial deepfakes were the exception in the political arena, reserved for narrow windows and contingent on plausibility or credible amplification, which raised risks for the amplifiers. Our analysis, moreover, can also explain why two other classes of adversarial representational deepfakes \textit{did} proliferate: sexual deepfakes and scam advertisements. Both of these have been observed widely in Taiwan, India, and globally. Clearly, both have increased in quality and scale with model improvements; both are bypassing distributional bottlenecks; and both are meeting some form of attentional demand. But in contrast with propaganda, neither is held back by creators' need to build a long-term reputation or distributors' fears of criminal liability---nudifiers operate within legally grey or unenforced areas~\citep{han2025characterizing, timmerman2023studying} and scammers are already engaging in criminal activity, so the use of AI does not create any additional marginal risk. The social context indicating the image's fakeness may actually be an asset, moreover, because they help separate the most gullible targets from those who are more discerning and thus might waste the scammer's time~\citep{herley2012nigerian}. It is thus socio-technical factors, rather than either technical or social constraints alone, that can explain why these two forms of deepfakes proliferated even as political deepfakes remained more limited.

\section{Interventions}

This work has three major implications for security researchers seeking to develop effective defenses. First, there is a need for researchers to be precise in how they define deepfakes as well as to look beyond deepfakes to consider the broader space of generative propaganda. Second, with respect to the problem of efficiency gains, we recommend measures to help scale defenses to meet the scale of the challenge. Finally, to defend against misuse by internal and participatory actors, we recommend developing mitigations that can complement existing social defenses. We provide our recommendations in three levels: model (AI models and software used to create content), platform (websites and channels through which the content is disseminated), and audience (the consumers of the content). Across levels, researchers will need to be cognizant of the social factors shaping both AI's use and the efficacy of interventions.

\subsection{Identify AI's Use} 

Section~\ref{sec:taxonomy} shows that observed uses were a superset of those that defenders expected because of the discursive power of the term ``deepfakes''. The security implications of this work are straightforward: defenses skewed towards the verification of authenticity will struggle with false positives for obvious, promotional representations, while under-prioritizing the narrative risk surface. Below, we present recommendations at each level of the stack given these findings. \\

\noindent \textbf{Model Level: Incorporate Visible/Audible Watermarks.} A watermark refers to the embedding of a perceptible or imperceptible message within digital content that identifies the content's source. Although there is no watermarking method that cannot easily be removed~\citep{zhao2024invisible, zhang2023watermarks} and technically savvy actors will be able to use their own watermark-free tooling, our research suggests that, if perceptible markers are included by default, many creators will choose to keep them. Moreover, the act of removing a watermark can indicate deceptive intent, which is useful in a legal context, while the proliferation of content with visible watermarks has the additional benefit of making AI's broadening use salient to more people, enhancing AI literacy.\\ 

\noindent \textbf{Platform Level: Scale and Contextualize AI Detection.} AIPasta operates at scale and across text, image, audio and video modes. Humans and algorithms alike perform poorly in distinguishing text written with AIPasta from  human-written text, enhancing its distortive potential~\citep{dash2025persuasive}. Defenders will need better tooling to detect the over-representation of AI-generated content in entire sets of comments or videos. But because simply tracking and making visible or downranking AI's use risks overstating false positives from promotional or obvious use, results should be contextualized with other digital signals as well as social context. For platforms, such tooling will be key in preventing the gaming of trending lists and algorithms; however, there is a particular need for the advancement of decentralized and scaled detection for contexts, as in Taiwan, where some major platforms cannot be assumed to be defensive contexts but rather could be neutral or even adversarial~\citep{niven2023evolution}. \\ 

\noindent \textbf{Audience Level: Clearly Differentiate Adversarial Deepfakes.} For journalists, researchers, and fact-checkers seeking to draw public attention to misuse, it is important to clearly differentiate deepfakes in order to avoid overindexing on false positives. The very choice to use the category and frame of ``deepfake'' encodes content with a certain techno-political power, disrupting the co-development of shared understanding and meaning-making among front-lines defenders and security researchers that is required for effective intervention. In the context of political communications, then, adversarial representational deepfakes can be differentiated from other representational uses by the \textit{hidden} nature of AI's use and the \textit{derogatory} nature of the depiction. Deepfakes, moreover, should be recognized as just one of a set of misuses that includes AIPasta, Precision Propaganda, and AI slop---with these other forms of use benefiting from key efficiency gains. By empirically taxonomizing a broad set of actual uses of generative AI in political communications, we seek to expand the set of uses and misuses that defenders track, and to improve the precision with which defenders can identify content in need of intervention (versus the shits and giggles). \\

\subsection{Counter Efficiency Gains}

Section~\ref{sec:motivations} highlights the extent to which AI's efficiency gains empower external actors and groups seeking to operate across linguistic and cultural barriers. For these actors, it is expanded capacities to evade detection, to operate across modes, and to operate across languages that require a concerted and coordinated defense. \\

\noindent \textbf{Model Level: Leverage Local Data for Alignment.} Model- and software guardrails can limit the ability of participatory and technically unskilled actors to wield not just deepfakes but also AIPasta effectively---as evidenced, for example, by the extent to which the AI images circulated during a recent border conflict in India included goriness at most via cartoon blood (consistent with image generators' guardrails against violent content) \citep{kava2025deepfakes}. But such guardrails can be brittle when context or language changes~\citep{deng2023multilingual}. A
solution lies in leveraging the insights of front-line defenders: incorporating fact-checker reports into red teaming operations, for example, can improve evaluation
across language and culture~\citep{cuevasvillalba2025anecdoctoring}. Such methods could and should be adapted to leverage crowd-sourced tiplines or community notes-style data sets that bridge across political divides depending on what sources hold the most legitimacy in a given place.  \\

\noindent \textbf{Platform Level: Develop Cross-Platform and Cross-Lingual Monitoring Methods.} External actors are most likely to benefit from large efficiency gains for their operations while being relatively less constrained by existing social defenses. But these actors are also known to use the same tactics across multiple places~\citep{goldsteinBenson2025AIpropaganda}, which means that successful defenses can also scale across places. 
Platform data is increasingly inaccessible to researchers and defenders alike, presenting a major need for methodological innovation. Promising efforts include tools that enable users to donate their data to researchers~\citep{garimella2024prevalent, zannettou2024analyzing}, automated auditing methods~\citep{haroon2022youtube, wsj2025insideTikTokAlgorithm}, and observatories in which consented participants provide ongoing screen access~\citep{feal2024introduction}. Importantly, such work can benefit from the same efficiency gains seen by adversaries, scaling across language and place to provide ongoing and comparative insight. \\ 

\noindent \textbf{Audience Level: Create Multilingual and Multimodal Media Literacy Content.} Content that teaches people about maintaining healthy media diets and inoculates them to likely distortive narratives has shown strong protective potential even across partisan divides~\citep{roozenbeek2022psychological, kozyreva2024toolbox}. Ensuring that such efforts match the scale of the challenge, however, will require defenders to reach audiences across video as well as text platforms, and across languages.  The technical need, then, is for tooling that can help people better engage online, for example by helping defenders more easily create content across languages and modes. To be most effective, moreover, such efforts should move beyond literacy to competency---a position advocated for, in particular, by Taiwanese defenders---actively teaching everyday people how to seed, spread, and amplify their own legitimate narratives, in recognition that everyone is responsible for the health of their digital public spheres.\\ 

\subsection{Reinforce Social Defenses}

Section~\ref{sec:threatmodel} highlights the extent to which social defenses helped keep political deepfakes in check in two major 2024 elections. Our recommendations, then, are to develop technical mitigations that bolster and maintain these safeguards. We again recommend investments in a toolkit of diverse interventions at all levels of the stack---from models to platforms to audiences---that can be co-created with local defenders, because even the most brilliant mitigation will fail if it is not suited to the established economic and political systems within which it is introduced. \\

\noindent \textbf{Model Level: Piggyback on Existing Systems for Verifying Users.} Developing methods for user verification can enhance the traceability of users and thus the enforceability of laws against misuse. For example, Indian factcheckers saw usage of the ElevenLabs software for voice clones drop off when it introduced a fee for usage, though the fee was negligible (\$1 per month), because it tied users to their (identified) payment accounts. In documenting the power of (salient, enforced) legal constraints on misuse, our work is aligned with~\citet{han2025characterizing} and \citet{timmerman2023studying}, who similarly show that members of sexual deepfake communities worry about legal repercussions. To the extent that differences in legality and enforcement help to explain differences in the usage of AI for political versus sexual deepfakes, our work also bolsters these researchers' calls for stronger legal defenses against non-consensual sexual deepfakes. We caution that such methods could also be abused for the oppression of speech, however, and thus there need to be careful legal and technical constraints on who is able to access such data, and under what conditions. \\

\noindent \textbf{Platform Level: Improve Reputation Indicators.} Reputational risk was a deterrent for content creators who had an identity (real or pseudonymous) at stake, but we lack reputation systems that effectively constrain faceless social media accounts operated by troll groups and content farms. Our findings thus highlight an opportunity to constrain the distribution of generative propaganda by improving reputational indicators for accounts that have at some point disseminated adversarial deepfakes or participated in AIPasta campaigns. Online reputation systems (e.g., feedback scores) are behavior ledgers: they collect, aggregate, and summarize past behaviors, informing prospective consumers of potential risks. However, reputation signals, particularly those with negative valence (e.g., dislikes), are often absent in social media platforms. Authenticity indicators (e.g., verified checkmarks) are more frequently available---but, when easily obtainable, can actually help dishonest users spread violating content, instead of constraining it~\citep{wired_musk_scammers_paradise}. Thus, we need signals that are robust to manipulation, hard for undeserving accounts to obtain, and that track the account's role in the dissemination of misleading content. Recent work on ``synthesized signals''---where researchers compute and display new interface signals based on an account's past behavior---demonstrate ways in which platforms could introduce indicators that satisfy these design goals~\citep{tsuchiya-p2p,synth-social-signals}. The development of such indicators also offers another opportunity to piggyback on existing social systems, such as efforts on the Taiwanese forum PTT to show user IP addresses as additional contextual information. Even in cases where platform administrators are unwilling or unable to carry out these changes, initiatives such as the Return YouTube Dislike add-on---which has 6M+ users~\citep{returnyoutubedislike2024}, demonstrate that external parties can introduce and benefit from reputation signals in online interfaces. \\  

\noindent \textbf{Audience Level: Empower Defenders with Certified Content.} Reducing the credibility of inauthentic content is as much about providing verified social context as it is about showing when content is false. That is, in addition to the current research on detecting when AI is used, there is a need for more ways to verify the provenance of legitimate content. An organization called Truepic, for example, has provided human rights defenders with a system that cryptographically binds images with certified metadata and preserves provenance through a verifiable chain of custody~\citep{ulbricht2022digital, truepic2024pai}. Such tooling directly meets a stated need of several journalists we interviewed, who asked for the ability to rapidly verify metadata so that they could better provide evidence during emerging events, when narratives were most subject to distortion. Adherence to an open standard, such as the \citet{c2pa2025spec22} specification, will be key to ensuring interoperability, so that journalists and citizen reporters can easily validate records across devices and platforms,  preempting distortion with accurate records of contested events. \\ 

\subsection{Limitations}

Our work is subject to a number of limitations. First, across both contexts, our interviewees were largely experts or elites; there is need for further research to better understand the experiences of audiences and targets of generative content. Second, the lead researchers on this project were external to the places studied and thus lacked the deep political, social, and historical context of local researchers; our identities and institutional affiliations also shaped the access we gained and the ways the interviews unfolded. We sought to counter these challenges by recruiting local researchers, by engaging in ongoing reflexive practices with members of the research team, and by member-checking key insights and quotes. Finally, our methods and sampling approach evolved between the two studies---from purely focusing on a well-mapped network of defenders in Taiwan to including both defenders and creators in India. Because we did not interview creators in Taiwan, then, some of our insights on AI's effects may over-index on the Indian context. Nevertheless, we believe the two contexts provide important and triangulating insights, and thus we present them both here.
\section{Conclusion}

This paper distinguishes adversarial deepfakes from other forms of generative AI's use and misuse in political speech. We identify deepfakes as just one of a much broader set of uses for generative AI and political communications. We further show the importance of legal factors, reputational economics, and social context in constraining the misuse of generative AI for deceptive and adversarial representations of candidates during two major elections. In contrast with threat models in which technical factors of model cost and quality constrain AI's use, which lead to predictions that deepfakes will proliferate in future elections, our work suggests that combined social and technical factors can largely keep such misuses in check. We point, however, to the efficiency gains AI offers key adversarial actors as having highly concerning potential to amplify the distortion of digital public spheres. To meet the challenge, security researchers should prioritize interventions that enable defenders to match the efficiency gains AI offers, including detection methods that scale and integrate with other sources of social context; monitoring methods to empower frontlines defenders across places; and traceability, reputation, and provenance mechanisms that bolster existing social defenses.

Defensive mechanisms that overindexed on English and Western contexts, failing to provide adequate defenses to low-resource and Global South contexts, have had major consequences for marginalized populations globally. If we are to avoid repeating the failures of the roll-outs of radio~\citep{thompson2007media} and social media~\citep{human2018report, zakrzewski2021-neglected-india} in the age of generative AI, we will need to build an effective defensive toolkit that can scale across places, and soon.

\ifCLASSOPTIONcompsoc
  \section*{Acknowledgments} We are grateful to Aparna Kalra for research assistance. This paper was improved by detailed feedback from Siddharth Suri, Mike Walker, Ben Cutler, Glen Weyl,  Josh Benaloh, and Eric Horvitz. The research was also shaped by insights from Neil Coles, Karen Easterbrook, Cormac Herley, Parker Bach, Dave Leichtman, and Vandinika Shukla. We received valuable guidance and context from Nick Monaco, Lev Nachman, Ai-Men Lau, Jiawai-Peter Cui, and Ankit Lal. We learned about the term ``soft fakes'' from Josh Lawson and ``AIPasta'' from Saloni Dash. Finally, we thank Whitney Hudson, Scott Counts, Katie Zoller, danah boyd, Nancy Baym, Weishung Liu, and Chris White for crucial support and guidance throughout this project. 
\else
  \section*{Acknowledgment}
\fi

\bibliographystyle{plainnat}
\bibliography{main}

\begin{thebibliography}{113}
\providecommand{\natexlab}[1]{#1}
\providecommand{\url}[1]{\texttt{#1}}
\expandafter\ifx\csname urlstyle\endcsname\relax
  \providecommand{\doi}[1]{doi: #1}\else
  \providecommand{\doi}{doi: \begingroup \urlstyle{rm}\Url}\fi

\bibitem[Abdullah et~al.(2024)Abdullah, Cheruvu, Kanchi, Chung, Gao, Jadliwala, and Viswanath]{abdullah2024analysis}
Sifat~Muhammad Abdullah, Aravind Cheruvu, Shravya Kanchi, Taejoong Chung, Peng Gao, Murtuza Jadliwala, and Bimal Viswanath.
\newblock An analysis of recent advances in deepfake image detection in an evolving threat landscape.
\newblock In \emph{2024 IEEE Symposium on Security and Privacy (SP)}, pages 91--109. IEEE, 2024.

\bibitem[Adnal(2024)]{adnal2024reasi}
Madhuri Adnal.
\newblock Why 'all eyes on reasi' is trending on social media?
\newblock Oneindia News, June 2024.
\newblock URL \url{https://www.oneindia.com/india/why-all-eyes-on-reasi-is-trending-on-social-media-3849455.html}.
\newblock Published June 11, 2024, 12:43 IST.

\bibitem[Akbar et~al.(2021)Akbar, Panda, Kukreti, Meena, and Pal]{akbar2021misinformation}
Syeda~Zainab Akbar, Anmol Panda, Divyanshu Kukreti, Azhagu Meena, and Joyojeet Pal.
\newblock Misinformation as a window into prejudice: Covid-19 and the information environment in india.
\newblock \emph{Proceedings of the ACM on human-computer interaction}, 4\penalty0 (CSCW3):\penalty0 1--28, 2021.

\bibitem[Barron(2025)]{barron2025taiwans}
Duncan Barron.
\newblock Taiwan’s model for digital democracy goes global.
\newblock \url{https://www.taipeitimes.com/News/feat/archives/2025/06/26/2003839259}, June 2025.
\newblock Taipei Times; Accessed: 2025-09-07.

\bibitem[Barry et~al.(1999)Barry, Britten, Barber, Bradley, and Stevenson]{barry1999using}
Christine~A Barry, Nicky Britten, Nick Barber, Colin Bradley, and Fiona Stevenson.
\newblock Using reflexivity to optimize teamwork in qualitative research.
\newblock \emph{Qualitative health research}, 9\penalty0 (1):\penalty0 26--44, 1999.

\bibitem[Baym(2015)]{baym2015personal}
Nancy~K Baym.
\newblock \emph{Personal connections in the digital age}.
\newblock John Wiley \& Sons, 2015.

\bibitem[Benkler et~al.(2018)Benkler, Faris, and Roberts]{benkler2018network}
Yochai Benkler, Robert Faris, and Hal Roberts.
\newblock \emph{Network propaganda: Manipulation, disinformation, and radicalization in American politics}.
\newblock Oxford University Press, 2018.

\bibitem[Bradshaw and Howard(2019)]{bradshaw2019global}
Samantha Bradshaw and Philip~N Howard.
\newblock The global disinformation order: 2019 global inventory of organised social media manipulation.
\newblock Technical report, Oxford Internet Institute, University of Oxford, 2019.
\newblock Published as part of the Computational Propaganda Project.

\bibitem[Budak et~al.(2024)Budak, Nyhan, Rothschild, Thorson, and Watts]{budak2024misunderstanding}
Ceren Budak, Brendan Nyhan, David~M Rothschild, Emily Thorson, and Duncan~J Watts.
\newblock Misunderstanding the harms of online misinformation.
\newblock \emph{Nature}, 630\penalty0 (8015):\penalty0 45--53, 2024.

\bibitem[{Burgess, Matt}(2022)]{wired_musk_scammers_paradise}
{Burgess, Matt}.
\newblock Elon musk's twitter is a scammer's paradise.
\newblock \emph{WIRED}, November 2022.
\newblock URL \url{https://www.wired.com/story/twitter-blue-check-verification-buy-scams/}.
\newblock Accessed: 2024-07-18.

\bibitem[Cao et~al.(2024)Cao, Das, Emami-Naeini, et~al.]{cao2024understanding}
Jiaxun Cao, Anupam Das, Pardis Emami-Naeini, et~al.
\newblock Understanding parents’ perceptions and practices toward children’s security and privacy in virtual reality.
\newblock In \emph{2024 IEEE Symposium on Security and Privacy (SP)}, pages 1554--1572. IEEE, 2024.

\bibitem[Chowdhury(2024)]{chowdhury2024AIConnectII5}
Rumman Chowdhury.
\newblock Keynote: Ai connect ii webinar 5 – ai explainability, accountability, and trust.
\newblock \url{https://www.atlanticcouncil.org/programs/geotech-center/ai-connect/ai-connect-ii-webinar-ai-explainability-accountability-and-trust/}, July 2024.
\newblock Atlantic Council GeoTech Center \& U.S. Department of State; Accessed: 2025-09-07.

\bibitem[Christopher(2024)]{christopher2024-instagram-ai-modi}
Nilesh Christopher.
\newblock Before india election, instagram boosts modi ai images that violate rules, April 2024.
\newblock URL \url{https://www.aljazeera.com/economy/2024/4/12/before-india-election-instagram-boosts-modi-ai-images-that-violate-rules}.

\bibitem[{Coalition for Content Provenance and Authenticity (C2PA)}(2025)]{c2pa2025spec22}
{Coalition for Content Provenance and Authenticity (C2PA)}.
\newblock Content credentials: {C2PA} technical specification, May 2025.
\newblock URL \url{https://spec.c2pa.org/specifications/specifications/2.2/index.html}.
\newblock Version 2.2 released 2025-05-01.

\bibitem[Cowen(2023)]{cowen2023}
Tyler Cowen.
\newblock Too much misinformation? the issue is demand, not supply.
\newblock {Bloomberg}, October 2023.
\newblock \url{https://www.bloomberg.com/opinion/articles/2023-10-03/campaign-2024-will-ai-generated-misinformation-be-a-big-problem}. Accessed 2025-07-01.

\bibitem[Cuevas et~al.(2025)Cuevas, Dash, Nayak, Vann, and Daepp]{cuevasvillalba2025anecdoctoring}
Alejandro Cuevas, Saloni Dash, Bharat~Kumar Nayak, Dan Vann, and Madeleine Daepp.
\newblock Anecdoctoring: Automated red-teaming across language and place.
\newblock In \emph{Proceedings of the Conference on Empirical Methods in Natural Language Processing (EMNLP)}. Association for Computational Linguistics, 2025.
\newblock Forthcoming.

\bibitem[Daepp and Ness(2024)]{daepp_ness_2024_video_will_kill_truth}
Madeleine I.~G. Daepp and Robert~Osazuwa Ness.
\newblock Video will kill the truth if monitoring doesn’t improve, argue two researchers.
\newblock \emph{The Economist}, 2024.
\newblock URL \url{https://www.economist.com/by-invitation/2024/03/26/video-will-kill-the-truth-if-monitoring-doesnt-improve-argue-two-researchers}.

\bibitem[Dash et~al.(2025)Dash, Xu, Jalbert, and Spiro]{dash2025persuasive}
Saloni Dash, Yiwei Xu, Madeline Jalbert, and Emma~S Spiro.
\newblock The persuasive potential of ai-paraphrased information at scale.
\newblock \emph{PNAS nexus}, 4\penalty0 (7):\penalty0 pgaf207, 2025.

\bibitem[{Deepfake Analysis Unit}(2024)]{dau2024-rajasthan-avatar-tamil}
{Deepfake Analysis Unit}.
\newblock A.i. avatar of politician from rajasthan speaks tamil, April 2024.
\newblock URL \url{https://dau.mcaindia.in/blog/a-i-avatar-of-politician-from-rajasthan-speaks-tamil}.
\newblock Published Apr 30, 2024; updated Jun 24, 2024.

\bibitem[Deng et~al.(2023)Deng, Zhang, Pan, and Bing]{deng2023multilingual}
Yue Deng, Wenxuan Zhang, Sinno~Jialin Pan, and Lidong Bing.
\newblock Multilingual jailbreak challenges in large language models.
\newblock \emph{arXiv preprint arXiv:2310.06474}, 2023.

\bibitem[Diel et~al.(2024)Diel, Lalgi, Schr{\"o}ter, MacDorman, Teufel, and B{\"a}uerle]{diel2024human}
Alexander Diel, Tania Lalgi, Isabel~Carolin Schr{\"o}ter, Karl~F MacDorman, Martin Teufel, and Alexander B{\"a}uerle.
\newblock Human performance in detecting deepfakes: A systematic review and meta-analysis of 56 papers.
\newblock \emph{Computers in Human Behavior Reports}, 16:\penalty0 100538, 2024.

\bibitem[Ellul(1965)]{ellul1965propaganda}
Jacques Ellul.
\newblock \emph{Propaganda: The formation of men's attitudes}.
\newblock Vintage, 1965.

\bibitem[{Factcheck Bureau}(2024)]{indiatoday2024sule}
{Factcheck Bureau}.
\newblock Supriya sule, nana patole bitcoin audios: What ai detection tools tell us.
\newblock \url{https://www.indiatoday.in/fact-check/story/supriya-sule-nana-patole-viral-audios-bitcoin-bjp-ai-detection-tools-deepfake-2636620-2024-11-20}, Nov 2024.
\newblock Accessed: 2025-09-06.

\bibitem[Feal et~al.(2024)Feal, Gleason, Goel, Radford, Yang, Basl, Meyer, Choffnes, Wilson, and Lazer]{feal2024introduction}
Alvaro Feal, Jeffrey Gleason, Pranav Goel, Jason Radford, Kai-Cheng Yang, John Basl, Michelle Meyer, David Choffnes, Christo Wilson, and David Lazer.
\newblock Introduction to national internet observatory.
\newblock In \emph{Workshop Proceedings of the 18th International AAAI Conference on Web and Social Media}, page~73, 2024.

\bibitem[{Focus Taiwan}(2023)]{focusTaiwan2023deepfakeJail}
{Focus Taiwan}.
\newblock Taiwan high court says deepfake porn producer must do jail time.
\newblock \url{https://focustaiwan.tw/society/202312280008}, December 2023.
\newblock Central News Agency English News. Accessed: 2025-09-07.

\bibitem[Frank et~al.(2024)Frank, Herbert, Ricker, Sch{\"o}nherr, Eisenhofer, Fischer, D{\"u}rmuth, and Holz]{frank2024representative}
Joel Frank, Franziska Herbert, Jonas Ricker, Lea Sch{\"o}nherr, Thorsten Eisenhofer, Asja Fischer, Markus D{\"u}rmuth, and Thorsten Holz.
\newblock A representative study on human detection of artificially generated media across countries.
\newblock In \emph{2024 IEEE Symposium on Security and Privacy (SP)}, pages 55--73. IEEE, 2024.

\bibitem[Freed et~al.(2019)Freed, Havron, Tseng, Gallardo, Chatterjee, Ristenpart, and Dell]{freed2019my}
Diana Freed, Sam Havron, Emily Tseng, Andrea Gallardo, Rahul Chatterjee, Thomas Ristenpart, and Nicola Dell.
\newblock " is my phone hacked?" analyzing clinical computer security interventions with survivors of intimate partner violence.
\newblock \emph{Proceedings of the ACM on Human-Computer Interaction}, 3\penalty0 (CSCW):\penalty0 1--24, 2019.

\bibitem[Gamage et~al.(2022)Gamage, Ghasiya, Bonagiri, Whiting, and Sasahara]{gamage2022deepfakes}
Dilrukshi Gamage, Piyush Ghasiya, Vamshi Bonagiri, Mark~E Whiting, and Kazutoshi Sasahara.
\newblock Are deepfakes concerning? analyzing conversations of deepfakes on reddit and exploring societal implications.
\newblock In \emph{Proceedings of the 2022 CHI conference on human factors in computing systems}, pages 1--19, 2022.

\bibitem[Garimella and Chauchard(2024)]{garimella2024prevalent}
Kiran Garimella and Simon Chauchard.
\newblock How prevalent is ai misinformation? what our studies in india show so far.
\newblock \emph{Nature}, 630\penalty0 (8015):\penalty0 32--34, 2024.

\bibitem[Geertz(2017)]{geertz2017interpretation}
Clifford Geertz.
\newblock \emph{The interpretation of cultures}.
\newblock Basic books, 2017.

\bibitem[Gibson et~al.(2025)Gibson, Olszewski, Brigham, Crowder, Butler, Traynor, Redmiles, and Kohno]{gibson2025analyzing}
Cassidy Gibson, Daniel Olszewski, Natalie~Grace Brigham, Anna Crowder, Kevin~RB Butler, Patrick Traynor, Elissa~M Redmiles, and Tadayoshi Kohno.
\newblock Analyzing the $\{$AI$\}$ nudification application ecosystem.
\newblock In \emph{34th USENIX Security Symposium (USENIX Security 25)}, pages 1--20, 2025.

\bibitem[Goldman(2024)]{deepfakedelugeventurebeat}
Sharon Goldman.
\newblock Deepfake deluge expected from ai image generation breakthrough (so long, lora?).
\newblock VentureBeat, January 2024.
\newblock \url{https://venturebeat.com/ai/deepfake-deluge-expected-from-ai-image-generation-breakthrough-so-long-lora/}. Accessed 2025-07-01.

\bibitem[Goldstein and Benson(2025)]{goldsteinBenson2025AIpropaganda}
Brett~J. Goldstein and Brett~V. Benson.
\newblock The era of a.i. propaganda has arrived, and america must act.
\newblock \url{https://www.nytimes.com/2025/08/05/opinion/china-ai-propaganda.html}, August 2025.
\newblock The New York Times. Accessed: 2025-09-07.

\bibitem[Guba(1981)]{guba1981criteria}
Egon~G Guba.
\newblock Criteria for assessing the trustworthiness of naturalistic inquiries.
\newblock \emph{Ectj}, 29\penalty0 (2):\penalty0 75--91, 1981.

\bibitem[Habermas(2023)]{habermas2023new}
J{\"u}rgen Habermas.
\newblock \emph{A new structural transformation of the public sphere and deliberative politics}.
\newblock John Wiley \& Sons, 2023.

\bibitem[Habgood-Coote(2023)]{habgood2023deepfakes}
Joshua Habgood-Coote.
\newblock Deepfakes and the epistemic apocalypse.
\newblock \emph{Synthese}, 201\penalty0 (3):\penalty0 103, 2023.

\bibitem[Han et~al.(2025)Han, Li, Kumar, and Durumeric]{han2025characterizing}
Catherine Han, Anne Li, Deepak Kumar, and Zakir Durumeric.
\newblock Characterizing the $\{$MrDeepFakes$\}$ sexual deepfake marketplace.
\newblock In \emph{34th USENIX Security Symposium (USENIX Security 25)}, pages 5169--5188, 2025.

\bibitem[Haroon et~al.(2022)Haroon, Chhabra, Liu, Mohapatra, Shafiq, and Wojcieszak]{haroon2022youtube}
Muhammad Haroon, Anshuman Chhabra, Xin Liu, Prasant Mohapatra, Zubair Shafiq, and Magdalena Wojcieszak.
\newblock Youtube, the great radicalizer? auditing and mitigating ideological biases in youtube recommendations.
\newblock \emph{arXiv preprint arXiv:2203.10666}, 2022.

\bibitem[Harvey(2015)]{harvey2015beyond}
Lou Harvey.
\newblock Beyond member-checking: A dialogic approach to the research interview.
\newblock \emph{International Journal of Research \& Method in Education}, 38\penalty0 (1):\penalty0 23--38, 2015.

\bibitem[Heidari et~al.(2024)Heidari, Jafari~Navimipour, Dag, and Unal]{heidari2024deepfake}
Arash Heidari, Nima Jafari~Navimipour, Hasan Dag, and Mehmet Unal.
\newblock Deepfake detection using deep learning methods: A systematic and comprehensive review.
\newblock \emph{Wiley Interdisciplinary Reviews: Data Mining and Knowledge Discovery}, 14\penalty0 (2):\penalty0 e1520, 2024.

\bibitem[Herley(2012)]{herley2012nigerian}
Cormac Herley.
\newblock Why do nigerian scammers say they are from nigeria?
\newblock In \emph{WEIS}. Berlin, 2012.

\bibitem[Hoffman(2024)]{hoffman2024first}
Benjamin Hoffman.
\newblock First came “spam.” now, with ai, we’ve got “slop.”.
\newblock \emph{The New York Times}, 2024.

\bibitem[Horvitz(2022)]{horvitz2022horizon}
Eric Horvitz.
\newblock On the horizon: Interactive and compositional deepfakes.
\newblock In \emph{proceedings of the 2022 international conference on multimodal interaction}, pages 653--661, 2022.

\bibitem[{Human Rights Council}(2018)]{human2018report}
{Human Rights Council}.
\newblock Report of the independent international fact-finding mission on myanmar.
\newblock \emph{United Nations}, 2018.

\bibitem[Im et~al.(2020)Im, Tandon, Chandrasekharan, Denby, and Gilbert]{synth-social-signals}
Jane Im, Sonali Tandon, Eshwar Chandrasekharan, Taylor Denby, and Eric Gilbert.
\newblock Synthesized social signals: Computationally-derived social signals from account histories.
\newblock In \emph{Proceedings of the 2020 CHI Conference on Human Factors in Computing Systems}, CHI '20, page 1–12, New York, NY, USA, 2020. Association for Computing Machinery.
\newblock ISBN 9781450367080.
\newblock \doi{10.1145/3313831.3376383}.
\newblock URL \url{https://doi.org/10.1145/3313831.3376383}.

\bibitem[Jack(2017)]{jack2017lexicon}
Caroline Jack.
\newblock Lexicon of lies: Terms for problematic information.
\newblock \emph{Data \& Society}, 3\penalty0 (22):\penalty0 1094--1096, 2017.

\bibitem[Kapoor and Narayanan(2023)]{kapoor2023}
Sayash Kapoor and Arvind Narayanan.
\newblock How to prepare for the deluge of generative ai on social media.
\newblock \emph{Knight First Amendment Institute Blog}, June 2023.
\newblock Available at: \url{https://knightcolumbia.org/content/how-to-prepare-for-the-deluge-of-generative-ai-on-social-media}. Accessed 2025-07-02.

\bibitem[Kapoor and Narayanan(2024)]{kapoor2024}
Sayash Kapoor and Arvind Narayanan.
\newblock We looked at 78 election deepfakes. political misinformation is not an ai problem.
\newblock \emph{Knight First Amendment Institute Blog}, December 2024.
\newblock Available at: \url{https://knightcolumbia.org/blog/we-looked-at-78-election-deepfakes-political-misinformation-is-not-an-ai-problem}. Accessed 2025-07-01.

\bibitem[Kava(2025)]{kava2025deepfakes}
Shivani Kava.
\newblock Deepfakes, voice clones and ai images amplified disinformation on india-pak conflict.
\newblock \emph{The News Minute}, June 2025.
\newblock URL \url{https://www.thenewsminute.com/news/deepfakes-voice-clones-and-ai-images-amplified-disinformation-on-india-pak-conflict}.
\newblock Edited by Bharathy Singaravel.

\bibitem[Kietzmann et~al.(2020)Kietzmann, Lee, McCarthy, and Kietzmann]{kietzmann2020deepfakes}
Jan Kietzmann, Linda~W Lee, Ian~P McCarthy, and Tim~C Kietzmann.
\newblock Deepfakes: Trick or treat?
\newblock \emph{Business horizons}, 63\penalty0 (2):\penalty0 135--146, 2020.

\bibitem[King et~al.(2017)King, Pan, and Roberts]{king2017chinese}
Gary King, Jennifer Pan, and Margaret~E Roberts.
\newblock How the chinese government fabricates social media posts for strategic distraction, not engaged argument.
\newblock \emph{American political science review}, 111\penalty0 (3):\penalty0 484--501, 2017.

\bibitem[Kozyreva et~al.(2024)Kozyreva, Lorenz-Spreen, Herzog, Ecker, Lewandowsky, Hertwig, Ali, Bak-Coleman, Barzilai, Basol, et~al.]{kozyreva2024toolbox}
Anastasia Kozyreva, Philipp Lorenz-Spreen, Stefan~M Herzog, Ullrich~KH Ecker, Stephan Lewandowsky, Ralph Hertwig, Ayesha Ali, Joe Bak-Coleman, Sarit Barzilai, Melisa Basol, et~al.
\newblock Toolbox of individual-level interventions against online misinformation.
\newblock \emph{Nature Human Behaviour}, 8\penalty0 (6):\penalty0 1044--1052, 2024.

\bibitem[Le et~al.(2025)Le, Kim, Woo, Moore, Abuadbba, and Tariq]{le2025sok}
Binh~M Le, Jiwon Kim, Simon~S Woo, Kristen Moore, Alsharif Abuadbba, and Shahroz Tariq.
\newblock Sok: Systematization and benchmarking of deepfake detectors in a unified framework.
\newblock In \emph{2025 IEEE 10th European Symposium on Security and Privacy (EuroS\&P)}, pages 883--902. IEEE, 2025.

\bibitem[Le~Blond et~al.(2018)Le~Blond, Cuevas, Troncoso-Pastoriza, Jovanovic, Ford, and Hubaux]{le2018enforcing}
Stevens Le~Blond, Alejandro Cuevas, Juan~Ram{\'o}n Troncoso-Pastoriza, Philipp Jovanovic, Bryan Ford, and Jean-Pierre Hubaux.
\newblock On enforcing the digital immunity of a large humanitarian organization.
\newblock In \emph{2018 IEEE Symposium on Security and Privacy (SP)}, pages 424--440. IEEE, 2018.

\bibitem[Lee(2020)]{lee2020nobody}
Mei-Chun Lee.
\newblock \emph{The “nobody” movement: Digital activism and the uprising of civic hackers in Taiwan}.
\newblock University of California, Davis, 2020.

\bibitem[Li(2024)]{li2024-seeing-not-believing-ii}
Wei-Ping Li.
\newblock Seeing is not believing (part ii) -- ai videos spread during the 2024 presidential election in taiwan, February 2024.
\newblock URL \url{https://en.tfc-taiwan.org.tw/en_tfc_294/}.

\bibitem[Lührmann et~al.(2019)Lührmann, Gastaldi, Grahn, Lindberg, Maxwell, Mechkova, Morgan, Stepanova, and Pillai]{vdem2019}
Anna Lührmann, Lisa Gastaldi, Sandra Grahn, Staffan~I. Lindberg, Laura Maxwell, Valeriya Mechkova, Richard Morgan, Natalia Stepanova, and Shreeya Pillai.
\newblock Varieties of democracy (v-dem) annual democracy report 2019: Democracy facing global challenges.
\newblock Technical report, {V-Dem Insitute}, 2019.

\bibitem[MacKenzie and Wajcman(1999)]{mackenzie1999social}
Donald MacKenzie and Judy Wajcman.
\newblock \emph{The social shaping of technology}.
\newblock Open University, 1999.

\bibitem[Matthews et~al.(2017)Matthews, O’Leary, Turner, Sleeper, Woelfer, Shelton, Manthorne, Churchill, and Consolvo]{matthews2017security}
Tara Matthews, Kathleen O’Leary, Anna Turner, Manya Sleeper, Jill~Palzkill Woelfer, Martin Shelton, Cori Manthorne, Elizabeth~F Churchill, and Sunny Consolvo.
\newblock Security and privacy experiences and practices of survivors of intimate partner abuse.
\newblock \emph{IEEE Security \& Privacy}, 15\penalty0 (5):\penalty0 76--81, 2017.

\bibitem[McGregor et~al.(2015)McGregor, Charters, Holliday, and Roesner]{mcgregor2015investigating}
Susan~E McGregor, Polina Charters, Tobin Holliday, and Franziska Roesner.
\newblock Investigating the computer security practices and needs of journalists.
\newblock In \emph{24th USENIX Security Symposium (USENIX Security 15)}, pages 399--414, 2015.

\bibitem[Meaker(2023)]{meaker2023slovakias}
Morgan Meaker.
\newblock Slovakia’s election deepfakes show {AI} is a danger to democracy.
\newblock \url{https://www.wired.com/story/slovakias-election-deepfakes-show-ai-is-a-danger-to-democracy/}, Oct 2023.
\newblock Accessed: 2025-09-06.

\bibitem[Mehta et~al.(2023)Mehta, Jagatap, Gallagher, Timmerman, Deb, Garg, Greenstadt, and Dolan-Gavitt]{mehta2023can}
Pulak Mehta, Gauri Jagatap, Kevin Gallagher, Brian Timmerman, Progga Deb, Siddharth Garg, Rachel Greenstadt, and Brendan Dolan-Gavitt.
\newblock Can deepfakes be created on a whim?
\newblock In \emph{Companion Proceedings of the ACM Web Conference 2023}, pages 1324--1334, 2023.

\bibitem[Meng et~al.(2024)Meng, Wang, Guo, Ju, and Zhao]{meng2024ava}
Xiangtao Meng, Li~Wang, Shanqing Guo, Lei Ju, and Qingchuan Zhao.
\newblock Ava: Inconspicuous attribute variation-based adversarial attack bypassing deepfake detection.
\newblock In \emph{2024 IEEE Symposium on Security and Privacy (SP)}, pages 74--90. IEEE, 2024.

\bibitem[Munyendo et~al.(2023)Munyendo, Acar, and Aviv]{munyendo2023eighty}
Collins~W Munyendo, Yasemin Acar, and Adam~J Aviv.
\newblock " in eighty percent of the cases, i select the password for them": Security and privacy challenges, advice, and opportunities at cybercafes in kenya.
\newblock In \emph{2023 IEEE Symposium on Security and Privacy (SP)}, pages 570--587. IEEE, 2023.

\bibitem[Munyendo et~al.(2025)Munyendo, Owens, Strong, Wang, Aviv, Kohno, and Roesner]{munyendo2025you}
Collins~W Munyendo, Kentrell Owens, Faith Strong, Shaoqi Wang, Adam~J Aviv, Tadayoshi Kohno, and Franziska Roesner.
\newblock “you have to ignore the dangers”: User perceptions of the security and privacy benefits of whatsapp mods.
\newblock In \emph{2025 IEEE Symposium on Security and Privacy (SP)}, pages 4515--4533. IEEE, 2025.

\bibitem[Niven(2023)]{niven2023evolution}
Tim Niven.
\newblock The evolution of china’s interference in taiwan.
\newblock \url{https://thediplomat.com/2023/12/the-evolution-of-chinas-interference-in-taiwan/}, December 2023.
\newblock The Diplomat; Accessed: 2025-09-07.

\bibitem[Paris and Donovan(2019)]{paris2019deepfakes}
Britt Paris and Joan Donovan.
\newblock Deepfakes and cheap fakes.
\newblock \emph{United States of America: Data \& Society}, 1, 2019.

\bibitem[Pawelec(2022)]{pawelec2022deepfakes}
Maria Pawelec.
\newblock Deepfakes and democracy (theory): How synthetic audio-visual media for disinformation and hate speech threaten core democratic functions.
\newblock \emph{Digital society}, 1\penalty0 (2):\penalty0 19, 2022.

\bibitem[Pearman et~al.(2019)Pearman, Zhang, Bauer, Christin, and Cranor]{pearman2019people}
Sarah Pearman, Shikun~Aerin Zhang, Lujo Bauer, Nicolas Christin, and Lorrie~Faith Cranor.
\newblock Why people (don't) use password managers effectively.
\newblock In \emph{Fifteenth symposium on usable privacy and security (SOUPS 2019)}, pages 319--338, 2019.

\bibitem[Pe{\~n}a-Alonso et~al.(2025)Pe{\~n}a-Alonso, Pe{\~n}a-Fern{\'a}ndez, and Meso-Ayerdi]{pena2025journalists}
Urko Pe{\~n}a-Alonso, Sim{\'o}n Pe{\~n}a-Fern{\'a}ndez, and Koldobika Meso-Ayerdi.
\newblock Journalists’ perceptions of artificial intelligence and disinformation risks.
\newblock \emph{Journalism and Media}, 6\penalty0 (3):\penalty0 133, 2025.

\bibitem[Ray et~al.(2021)Ray, Wolf, Kuber, and Aviv]{ray2021older}
Hirak Ray, Flynn Wolf, Ravi Kuber, and Adam~J Aviv.
\newblock Why older adults (don't) use password managers.
\newblock In \emph{30th USENIX Security Symposium (USENIX Security 21)}, pages 73--90, 2021.

\bibitem[Redmiles et~al.(2016)Redmiles, Malone, and Mazurek]{redmiles2016think}
Elissa~M Redmiles, Amelia~R Malone, and Michelle~L Mazurek.
\newblock I think they're trying to tell me something: Advice sources and selection for digital security.
\newblock In \emph{2016 IEEE Symposium on Security and Privacy (SP)}, pages 272--288. IEEE, 2016.

\bibitem[Reed(2013)]{reed2013power}
Isaac~Ariail Reed.
\newblock Power: Relational, discursive, and performative dimensions.
\newblock \emph{Sociological Theory}, 31\penalty0 (3):\penalty0 193--218, 2013.

\bibitem[{Return YouTube Dislike}(2024)]{returnyoutubedislike2024}
{Return YouTube Dislike}.
\newblock Return youtube dislike ― chrome web store.
\newblock \url{https://chromewebstore.google.com/detail/return-youtube-dislike/gebbhagfogifgggkldgodflihgfeippi}, October 2024.
\newblock Accessed: 2025-09-15.

\bibitem[Reyes(2006)]{reyes2006swift}
G~Mitchell Reyes.
\newblock The swift boat veterans for truth, the politics of realism, and the manipulation of vietnam remembrance in the 2004 presidential election.
\newblock \emph{Rhetoric and Public Affairs}, pages 571--600, 2006.

\bibitem[Richet(2022)]{richet2022cybercriminal}
Jean-Loup Richet.
\newblock How cybercriminal communities grow and change: An investigation of ad-fraud communities.
\newblock \emph{Technological Forecasting and Social Change}, 174:\penalty0 121282, 2022.

\bibitem[Rifkind(2024)]{deepfakedelugetimes}
Hugo Rifkind.
\newblock The deepfake deluge has only just begun.
\newblock The Times, January 2024.
\newblock \url{https://www.thetimes.com/article/the-deepfake-deluge-has-only-just-begun-cq82q5zrb}. Accessed 2025-07-01.

\bibitem[Roozenbeek et~al.(2022)Roozenbeek, Van Der~Linden, Goldberg, Rathje, and Lewandowsky]{roozenbeek2022psychological}
Jon Roozenbeek, Sander Van Der~Linden, Beth Goldberg, Steve Rathje, and Stephan Lewandowsky.
\newblock Psychological inoculation improves resilience against misinformation on social media.
\newblock \emph{Science advances}, 8\penalty0 (34):\penalty0 eabo6254, 2022.

\bibitem[Roy(2025)]{nytimes2025riseAIinfluencer}
Jessica Roy.
\newblock The rise of the ai influencer.
\newblock \url{https://www.nytimes.com/2025/09/03/style/ai-influencers-lil-miquela-mia-zelu.html}, September 2025.
\newblock The New York Times. Accessed: 2025-09-06.

\bibitem[Saldana(2011)]{saldana2011fundamentals}
Johnny Saldana.
\newblock \emph{Fundamentals of qualitative research}.
\newblock Oxford university press, 2011.

\bibitem[Schneier and Sanders(2024)]{schneier2024review}
Bruce Schneier and Nathan Sanders.
\newblock The apocalypse that wasnt: Ai was everywhere in 2024's elections, but deepfakes and misinformation were only part of the picture.
\newblock {The Conversation}, December 2024.
\newblock Available at: \url{https://theconversation.com/the-apocalypse-that-wasnt-ai-was-everywhere-in-2024s-elections-but-deepfakes-and-misinformation-were-only-part-of-the-picture-244225}. Accessed 2025-07-01.

\bibitem[Shan(2023)]{shan2023deepfakeFraud}
Shelley Shan.
\newblock Legislature passes stiffer jail, fine for deepfake fraud.
\newblock \url{https://www.taipeitimes.com/News/front/archives/2023/05/17/2003799936}, May 2023.
\newblock Taipei Times; Accessed: 2025-09-07.

\bibitem[Silverman(2018)]{silverman2018-spot-deepfake}
Craig Silverman.
\newblock How to spot a {DeepFake} like the barack obama--jordan peele video.
\newblock \emph{BuzzFeed News}, April 2018.
\newblock URL \url{https://www.buzzfeed.com/craigsilverman/obama-jordanpeele-deepfake-video-debunk-buzzfeed}.

\bibitem[Simon et~al.(2023)Simon, Altay, and Mercier]{simon2023misinformation}
Felix~M Simon, Sacha Altay, and Hugo Mercier.
\newblock Misinformation reloaded? fears about the impact of generative ai on misinformation are overblown.
\newblock \emph{Harvard Kennedy School Misinformation Review}, 4\penalty0 (5), 2023.

\bibitem[Small and Calarco(2022)]{small2022qualitative}
Mario~Luis Small and Jessica~McCrory Calarco.
\newblock \emph{Qualitative literacy: A guide to evaluating ethnographic and interview research}.
\newblock Univ of California Press, 2022.

\bibitem[Sorkin et~al.(2023)Sorkin, Warner, Kessler, de~la Merced, Hirsch, and Livni]{nyt-2023-ai-spoof-markets}
Andrew~Ross Sorkin, Bernhard Warner, Sarah Kessler, Michael~J. de~la Merced, Lauren Hirsch, and Ephrat Livni.
\newblock An a.i.-generated spoof rattles the markets.
\newblock \emph{The New York Times}, May 2023.
\newblock URL \url{https://www.nytimes.com/2023/05/23/business/ai-picture-stock-market.html}.

\bibitem[Starbird(2024)]{starbird2024facts}
K~Starbird.
\newblock Facts, frames, and (mis) interpretations: understanding rumors as collective sensemaking.
\newblock \emph{Center for an Informed Public https://www. cip. uw. edu/2023/12/06/rumors-collective-sensemaking-kate-starbird/(Univ. Washington, 2023)}, 2024.

\bibitem[Starbird et~al.(2019)Starbird, Arif, and Wilson]{starbird2019disinformation}
Kate Starbird, Ahmer Arif, and Tom Wilson.
\newblock Disinformation as collaborative work: Surfacing the participatory nature of strategic information operations.
\newblock \emph{Proceedings of the ACM on human-computer interaction}, 3\penalty0 (CSCW):\penalty0 1--26, 2019.

\bibitem[{Taiwan FactCheck Center}(2023)]{tfc-2024-election-ai-audio-2023}
{Taiwan FactCheck Center}.
\newblock 2024 election fact-checking notes, episode 1: Taiwan's first pre-election ai-fabricated audio; tips to identify fake audio and video, October 2023.
\newblock URL \url{https://tfc-taiwan.org.tw/migration_article_105929_9781/}.
\newblock Original in Chinese. Last accessed 2025-09-17.

\bibitem[{Taiwan FactCheck Center}(2024)]{tfc_10288_robot_farming_2024}
{Taiwan FactCheck Center}.
\newblock [manipulated video] viral clips claim that with today's advanced technology, robots can replace farm labor — is it true?
\newblock Taiwan FactCheck Center, February 2024.
\newblock URL \url{https://tfc-taiwan.org.tw/articles/10288}.
\newblock Fact-Check Report No. 2851. Published 2024-02-06. Last accessed 2025-09-15.

\bibitem[{The Economist}(2023)]{economistelections}
{The Economist}.
\newblock 2024 is the biggest election year in history.
\newblock The Economist, November 2023.
\newblock \url{https://www.economist.com/interactive/the-world-ahead/2023/11/13/2024-is-the-biggest-election-year-in-history}. Accessed 2025-06-26.

\bibitem[{The Wall Street Journal}(2021)]{wsj2025insideTikTokAlgorithm}
{The Wall Street Journal}.
\newblock Inside tiktok’s highly secretive algorithm.
\newblock \url{https://www.wsj.com/video/series/inside-tiktoks-highly-secretive-algorithm}, 2021.
\newblock The Wall Street Journal. Accessed: 2025-09-07.

\bibitem[Thompson(2007)]{thompson2007media}
Allan Thompson.
\newblock \emph{The media and the Rwanda genocide}.
\newblock IDRC, 2007.

\bibitem[Tilley and Woodthorpe(2011)]{tilley2011end}
Liz Tilley and Kate Woodthorpe.
\newblock Is it the end for anonymity as we know it? a critical examination of the ethical principle of anonymity in the context of 21st century demands on the qualitative researcher.
\newblock \emph{Qualitative research}, 11\penalty0 (2):\penalty0 197--212, 2011.

\bibitem[Timmerman et~al.(2023)Timmerman, Mehta, Deb, Gallagher, Dolan-Gavitt, Garg, and Greenstadt]{timmerman2023studying}
Brian Timmerman, Pulak Mehta, Progga Deb, Kevin Gallagher, Brendan Dolan-Gavitt, Siddharth Garg, and Rachel Greenstadt.
\newblock Studying the online deepfake community.
\newblock \emph{Journal of Online Trust and Safety}, 2\penalty0 (1), 2023.

\bibitem[Tran et~al.(2024)Tran, Munyendo, Ramulu, Rodriguez, Schnell, Sula, Simko, and Acar]{tran2024security}
Mindy Tran, Collins~W Munyendo, Harshini~Sri Ramulu, Rachel~Gonzalez Rodriguez, Luisa~Ball Schnell, Cora Sula, Lucy Simko, and Yasemin Acar.
\newblock Security, privacy, and data-sharing trade-offs when moving to the united states: Insights from a qualitative study.
\newblock In \emph{2024 IEEE Symposium on Security and Privacy (SP)}, pages 617--634. IEEE, 2024.

\bibitem[Trauthig et~al.(2025)Trauthig, Howard, and S]{trauthig2025}
I~Trauthig, PN~Howard, and Valenzuela S.
\newblock The role of generative ai use in 2024 elections worldwide, 2025.
\newblock Technical Paper, TP2025.2, doi: 10.61452/HZUE9853. Accessed: 2025-06-30.

\bibitem[Travelli(2024)]{travelli2024nyt}
Alex Travelli.
\newblock A small army combating a flood of deepfakes in india's election.
\newblock \url{https://www.nytimes.com/2024/06/01/world/asia/india-election-deepfakes.html}, June 2024.
\newblock The New York Times. Accessed: 2025-09-07.

\bibitem[{Truepic}(2024)]{truepic2024pai}
{Truepic}.
\newblock How truepic used disclosures to help authenticate cultural heritage imagery in conflict zones.
\newblock \url{https://partnershiponai.org/wp-content/uploads/2024/11/case-study-truepic.pdf}, November 2024.
\newblock Accessed 9=7-2025.

\bibitem[Tsai and Chin(2024)]{tsai-chin2024-secret-history}
Yung-yao Tsai and Jonathan Chin.
\newblock China is posting fake videos of president: sources, January 2024.
\newblock URL \url{https://www.taipeitimes.com/News/front/archives/2024/01/11/2003811930}.
\newblock Images reproduced with permission from Dr. Austin Wang.

\bibitem[Tsuchiya et~al.(2024)Tsuchiya, Cuevas, and Christin]{tsuchiya-p2p}
Taro Tsuchiya, Alejandro Cuevas, and Nicolas Christin.
\newblock Identifying risky vendors in cryptocurrency p2p marketplaces.
\newblock In \emph{Proceedings of the ACM Web Conference}, WWW '24, page 99–110. Association for Computing Machinery, 2024.
\newblock \doi{10.1145/3589334.3645475}.

\bibitem[Twomey et~al.(2023)Twomey, Ching, Aylett, Quayle, Linehan, and Murphy]{twomey2023deepfake}
John Twomey, Didier Ching, Matthew~Peter Aylett, Michael Quayle, Conor Linehan, and Gillian Murphy.
\newblock Do deepfake videos undermine our epistemic trust? a thematic analysis of tweets that discuss deepfakes in the russian invasion of ukraine.
\newblock \emph{Plos one}, 18\penalty0 (10):\penalty0 e0291668, 2023.

\bibitem[Ulbricht et~al.(2022)Ulbricht, Moxley, Austin, and Norburg]{ulbricht2022digital}
Bailey~R Ulbricht, Christopher Moxley, Mackenzie~D Austin, and Molly~D Norburg.
\newblock Digital eyewitnesses: Using new technologies to authenticate evidence in human rights litigation.
\newblock \emph{Stan. L. Rev.}, 74:\penalty0 851, 2022.

\bibitem[Usman et~al.(2025)Usman, Sadik, Elgedawy, Ruoti, Zappala, et~al.]{usman2025security}
Warda Usman, John Sadik, Ran Elgedawy, Scott Ruoti, Daniel Zappala, et~al.
\newblock Security and privacy experiences of first-and second-generation pakistani immigrants to the us: Perceptions, practices, challenges, and parent-child dynamics.
\newblock In \emph{2025 IEEE Symposium on Security and Privacy (SP)}, pages 2284--2302. IEEE, 2025.

\bibitem[Widder et~al.(2022)Widder, Nafus, Dabbish, and Herbsleb]{widder2022limits}
David~Gray Widder, Dawn Nafus, Laura Dabbish, and James Herbsleb.
\newblock Limits and possibilities for “ethical ai” in open source: A study of deepfakes.
\newblock In \emph{Proceedings of the 2022 ACM Conference on Fairness, Accountability, and Transparency}, pages 2035--2046, 2022.

\bibitem[Woolley and Howard(2018)]{woolley2018computational}
Samuel~C Woolley and Philip~N Howard.
\newblock \emph{Computational propaganda: Political parties, politicians, and political manipulation on social media}.
\newblock Oxford University Press, 2018.

\bibitem[{World Economic Forum}(2024)]{wefrisks2024}
{World Economic Forum}.
\newblock The global risks report 2024, 19th edition, January 2024.
\newblock \url{https://www3.weforum.org/docs/WEF_The_Global_Risks_Report_2024.pdf}. Accessed 2025-07-01.

\bibitem[{X}(2024)]{x_soldiersaffron7_clip}
{X}.
\newblock Post by @soldiersaffron7.
\newblock \url{https://x.com/SoldierSaffron7/status/1790610601649319940}, 2024.
\newblock Social media post.

\bibitem[Yu et~al.(2025)Yu, Sharma, Hu, Wang, and Wang]{yu2025exploring}
Yaman Yu, Tanusree Sharma, Melinda Hu, Justin Wang, and Yang Wang.
\newblock Exploring parent-child perceptions on safety in generative ai: concerns, mitigation strategies, and design implications.
\newblock In \emph{2025 IEEE Symposium on Security and Privacy (SP)}, pages 2735--2752. IEEE, 2025.

\bibitem[Zakrzewski et~al.(2021)Zakrzewski, Vynck, Masih, and Mahtani]{zakrzewski2021-neglected-india}
Cat Zakrzewski, Gerrit~De Vynck, Niha Masih, and Shibani Mahtani.
\newblock {How Facebook neglected the rest of the world, fueling hate speech and violence in India}.
\newblock \emph{The Washington Post}, October 2021.
\newblock URL \url{https://www.washingtonpost.com/technology/2021/10/24/india-facebook-misinformation-hate-speech/}.
\newblock Accessed September 7, 2025.

\bibitem[Zannettou et~al.(2024)Zannettou, Nemes-Nemeth, Ayalon, Goetzen, Gummadi, Redmiles, and Roesner]{zannettou2024analyzing}
Savvas Zannettou, Olivia Nemes-Nemeth, Oshrat Ayalon, Angelica Goetzen, Krishna~P Gummadi, Elissa~M Redmiles, and Franziska Roesner.
\newblock Analyzing user engagement with tiktok's short format video recommendations using data donations.
\newblock In \emph{Proceedings of the 2024 CHI Conference on Human Factors in Computing Systems}, pages 1--16, 2024.

\bibitem[Zhang et~al.(2023)Zhang, Edelman, Francati, Venturi, Ateniese, and Barak]{zhang2023watermarks}
Hanlin Zhang, Benjamin~L Edelman, Danilo Francati, Daniele Venturi, Giuseppe Ateniese, and Boaz Barak.
\newblock Watermarks in the sand: Impossibility of strong watermarking for generative models.
\newblock \emph{arXiv preprint arXiv:2311.04378}, 2023.

\bibitem[Zhao et~al.(2024)Zhao, Zhang, Su, Vasan, Grishchenko, Kruegel, Vigna, Wang, and Li]{zhao2024invisible}
Xuandong Zhao, Kexun Zhang, Zihao Su, Saastha Vasan, Ilya Grishchenko, Christopher Kruegel, Giovanni Vigna, Yu-Xiang Wang, and Lei Li.
\newblock Invisible image watermarks are provably removable using generative ai.
\newblock \emph{Advances in neural information processing systems}, 37:\penalty0 8643--8672, 2024.

\end{thebibliography}

\end{document}